\documentclass[prd,twocolumn,superscriptaddress,amssymb,floatfix,nofootinbib]{revtex4-2}

\usepackage{amsfonts}
\usepackage[colorlinks=true,citecolor=blue,linkcolor=blue,urlcolor=blue]{hyperref}
\usepackage{color,xcolor}
\usepackage{graphicx,multirow}
\usepackage{amsmath}
\usepackage{color,amsmath,amssymb,graphicx,latexsym,subfigure}
\usepackage{threeparttable,multirow,makecell,tabularx}
\usepackage{ulem}

\newcommand\add[1]{{\color{blue}\uwave{#1}}}

\allowdisplaybreaks

\begin{document}

\title{Probing the $\gamma\gamma^*\to \eta^{(\prime)}$ Transition Form Factors with Newly Derived $\eta^{(\prime)}$-Meson Light-Cone Distribution Amplitudes}

\author{Dan-Dan Hu}
\email{hudd@stu.cqu.edu.cn}
\author{Xing-Gang Wu}
\email{wuxg@cqu.edu.cn}
\author{Yu-Jie Zhang}
\email{zhangyj@stu.cqu.edu.cn}
\address{Department of Physics, Chongqing Key Laboratory for Strongly Coupled Physics, Chongqing University, Chongqing 401331, P.R. China}
\author{Hai-Bing Fu}
\email{fuhb@gzmu.edu.cn (Corresponding author)}
\author{Tao Zhong}
\email{zhongtao@gzmu.edu.cn}
\address{Department of Physics, Guizhou Minzu University, Guiyang 550025, P.R. China}

\date{\today}

\begin{abstract}

In the present work, we analyze the properties of the transition form factors (TFFs) for the $\gamma\gamma^*\to \eta^{(\prime)}$ process, employing the $\eta^{(\prime)}$-meson light-cone distribution amplitude (LCDA) derived within the light-cone sum rule framework. To this end, we adopt the quark-flavor mixing scheme for the $\eta^{(\prime)}$ meson, and compute the TFFs by systematically incorporating transverse-momentum corrections and contributions beyond the leading Fock state. We utilize light-cone harmonic oscillator models to parameterize the longitudinal and transverse behavior of the leading-twist light-cone wavefunction, for which the corresponding LCDA exhibits a unimodal profile. We further examine the potential contributions of intrinsic charm components to the scaled TFFs $Q^2 F_{\eta\gamma}(Q^2)$ and $Q^2 F_{\eta'\gamma}(Q^2)$. Leveraging a range of values for the decay constant $f_{\eta_{c_0}}$ and implementing the $\eta$–$\eta'$–$\eta_c$ and $\eta$–$\eta'$–$G$-$\eta_c$ mixing mechanisms accordingly, together with the recently updated mixing angles, we investigate the impact of the intrinsic $c\bar{c}$ and gluonic component on these observables. In high-$Q^2$ regime, $Q^2 F_{\eta'\gamma}(Q^2)$ exhibits a marked increase in sensitivity to the charm quark component, whereas $Q^2F_{\eta\gamma}(Q^2)$ becomes notably stabilized. A detailed discussion of $\chi^2/d.o.f$ and $p$-values indicates that the intrinsic charm quark component is important and yields a substantial, non-negligible contribution across the entire $Q^2$ range.

\end{abstract}

\date{\today}

\maketitle

\section{Introduction}

Hadronic transition form factors (TFFs) are among the most critical physical observables for describing the mechanisms of exclusive reactions. Studying hadronic TFFs not only yields valuable insights into the internal structure of hadrons but also deepens our understanding of the transition from perturbative to non-perturbative behavior in the intermediate energy regime. In particular, the photon-meson TFFs of neutral pseudoscalar mesons, such as $\pi^0$, $\eta$, and $\eta$, have garnered broad interest due to their intimate connection with non-perturbative strong interaction dynamics~\cite{Braaten:1982yp}. Investigating these TFFs poses a multifaceted challenge, requiring a coherent integration of experimental, theoretical, and phenomenological approaches to fully characterize and predict their behavior. Furthermore, these TFFs serve as a valuable testing ground for precision studies of the Standard Model (SM), particularly in the context of Quantum Chromodynamics (QCD). Moreover, the study of flavor-singlet pseudoscalar mesons (e.g., $\eta$ and $\eta'$) is a highly active research area, offering a vital probe for physics beyond the SM~\cite{Gan:2020aco}.

Current research on the production mechanisms of $\eta$ and $\eta'$ mesons in hard processes remains relatively limited. Unlike pions, $\eta$ and $\eta'$ mesons may contain significant gluonium (glueball) components or intrinsic charm. This characteristic presents challenges for traditional approaches based on low-energy effective field theories when describing such hard processes -- given that these processes are highly sensitive to the behavior of meson wavefunctions at small quark separations. While several studies~\cite{Feldmann:1998vh, Kroll:2002nt, Huang:2006as, Kroll:2012gsh, Agaev:2014wna} have explored this possibility using various methods, no definitive conclusions have been reached regarding the existence or precise proportion of these exotic components. In recent years, the photon-meson TFFs for neutral pseudoscalar mesons have garnered significant attention, owing to their role in SM quantum corrections to ${(g-2)}_\mu$, the anomalous magnetic moment of the muon~\cite{Colangelo:2019uex, Aoyama:2020ynm, Chao:2020kwq, Burri:2022gdg, Estrada:2024cfy}. Early experimental investigations into the $\gamma\gamma^*\to \eta^{(\prime)}$ TFFs date back to measurements by the PLUTO collaboration in 1984~\cite{PLUTO:1984yyu}. Subsequently, the CLEO, L3, CELLO, and BABAR collaborations also measured the TFFs for $\gamma\gamma^*\to \eta^{(\prime)}$: these measurements were performed at space-like momentum transfers in the range $1.5-40~{\rm GeV^2}$~\cite{TPCTwoGamma:1990dho, CELLO:1990klc, L3:1997ocz, CLEO:1997fho, BaBar:2011nrp}, as well as at an exceptionally large time-like momentum transfer of $112~{\rm GeV^2}$ for $\gamma^*\to \eta^{(\prime)}\gamma$ TFFs~\cite{BaBar:2006ash}.

Theoretically, several groups have investigated the $\gamma\gamma^*\to \eta^{(\prime)}$ process via a variety of approaches, including the rational approximants~\cite{Escribano:2013kba, Escribano:2015nra}, the anomaly sum rule~\cite{Klopot:2012hd}, the chiral perturbation theory ($\chi$PT)~\cite{Burri:2022gdg, Czyz:2012nq}, the light-front holographic QCD~\cite{Swarnkar:2015osa}, the light-front quark model (LFQM)~\cite{Choi:2017zxn}, the Nambu-Jona-Lasinio (NJL) model~\cite{GomezDumm:2016bxp, Noguera:2011fv}, and the continuum approach using the Bethe-Salpeter (BS) wavefunction~\cite{Ding:2018xwy}, and etc.. These methods are typically formulated in terms of the flavor-singlet $\eta_1$ and flavor-octet $\eta_8$ states, with the introduction of two mixing angles $\theta_1$ and $\theta_8$~\cite{Cao:1999fs, Cao:2012nj, Bass:2018xmz}; and some works also incorporate gluon components~\cite{Agaev:2003kb, Kroll:2002nt, Kroll:2012gsh, Agaev:2014wna}, which elevates the complexity of the associated calculations. While discrepancies exist among the results yielded by these theoretical frameworks, the majority of predictions lie within the error margins of experimental measurements.

In addition to the flavor singlet and octet schemes employed for calculating the  $\gamma\gamma^*\to \eta^{(\prime)}$ TFFs, the quark flavor mixing scheme has also been utilized. A key advantage of this mixing scheme is that it not only simplifies calculations by involving only a single mixing angle $\phi$, but is also favored by theoretical and phenomenological considerations within the quark flavor basis~\cite{Feldmann:1997vc, Leutwyler:1997yr}. In the high-energy regime, TFFs can be computed asymptotically at leading twist by convolving the perturbative hard-scattering amplitude with the light meson's light-cone distribution amplitudes (LCDAs), thereby incorporating non-perturbative quantum chromodynamics (QCD) dynamics~\cite{Lepage:1980fj, Brodsky:1981rp, Chernyak:1983ej}. Using the light-cone perturbative QCD (pQCD) approach, one can derive an expression for photon-meson TFFs that retains the transverse momentum ($k_\perp$) dependence in both the hard-scattering amplitude and the light-cone wavefunction (LCWF)~\cite{Cao:1996di, Lepage:1980fj}. These LCWFs, in turn, are closely related to non-perturbative LCDAs. However, in contrast to the steadily improving precision of perturbative QCD hard amplitudes, non-perturbative LCDAs remain insufficiently characterized. Hadronic LCDAs provide a key window into the QCD interactions of quarks, antiquarks, and gluons within hadrons, thus serving as essential ingredients for applying QCD to hard exclusive processes. As such, the significance of hadronic LCDAs in predicting TFFs across both non-perturbative and perturbative momentum transfer regimes is widely recognized~\cite{Choi:2007yu, Efremov:1979qk}.

The pion LCDAs have been extensively investigated, cf. recent progresses~\cite{Zhong:2021epq, Dalley:2002nj, Praszalowicz:2001pi, Gao:2022vyh, LatticeParton:2022zqc}, and the pion-photon TFF has been computed up to next-to-next-to-leading order (NNLO) QCD corrections, with results from various methodologies showing good consistency~\cite{Gao:2021iqq, Melic:2002ij, Zhou:2023ivj, Braun:2021grd, Mondal:2021czk}. In contrast, the phenomenology of $\eta$ and $\eta^\prime$ mesons is considerably more complex due to their flavor mixing, rendering their study more challenging than that of pion. As a result, the quark-gluon structure of these mesons and the hard exclusive processes involving them have become the focus of numerous theoretical efforts. Current theoretical investigations into the LCDAs of $\eta^{(\prime)}$ mesons primarily employ an expansion framework based on Gegenbauer moments. In prior works, two typical approaches have been adopted to determine these moments for the leading-twist LCDAs: one derives them via methods analogous to those applied to the pion~\cite{Ball:2004ye, Ball:2007hb, Duplancic:2015zna}, while the other estimates the moments by fitting to experimental data~\cite{CLEO:1997fho, BaBar:2011nrp, Kroll:2012gsh}. Despite these advances, research on the LCDAs of $\eta^{(\prime)}$ mesons remains relatively limited to date.

It is well established that LCDAs, universal non-perturbative quantities, characterize the momentum-fraction distributions of constituents within a meson for a specific Fock state. These distribution amplitudes enter the calculation of exclusive processes via proper factorization theorems, thereby serving as pivotal inputs for QCD predictions of such processes. Moreover, a close correlation exists between the LCWF and the LCDA, e.g. the LCDA is obtained by integrating out the transverse momentum of the LCWF. For the $\eta^{(\prime)}$-meson LCWF, the Brodsky-Huang-Lepage (BHL) prescription~\cite{bhl} is typically employed to construct it. This prescription connects the equal-time wavefunction in the meson rest frame to the wavefunction in the infinite momentum frame. By properly tuning the model parameters, distinct asymptotic behaviors of $\eta^{(\prime)}$-meson LCDAs can be realized, ranging from asymptotic-like to Chernyak-Zhitnitsky~\cite{Chernyak:1981zz} (CZ)-like profiles, cf. Refs.\cite{Wu:2011gf, Huang:2006as}. This flexibility enables the identification of the most appropriate LCDA behavior for explaining experimental form factor data, though this objective has not yet been accomplished via dedicated quantitative calculations.

Building upon these theoretical foundations and the current understanding of LCDAs for $\eta^{(\prime)}$ mesons, we have carried out detailed calculations of the leading-twist LCDA of $\eta^{(\prime)}$ mesons using QCD sum rules within the background field theory (BFT) framework~\cite{Hu:2021zmy, Hu:2023pdl}. These calculations yield deeper insights into the internal structure of $\eta^{(\prime)}$ mesons and their momentum-fraction distributions. Such theoretical advancements pave the way for more precise comparisons between experimental measurements and theoretical predictions, ultimately advancing our understanding of meson decay dynamics within the QCD framework.

The leading-twist LCDAs of $\eta^{(\prime)}$ mesons are essential for interpreting both their corresponding semileptonic decays and the $\gamma\gamma^*\to\eta^{(\prime)}$ TFFs. Drawing on prior works~\cite{Hu:2021zmy, Hu:2023pdl, Zhang:2025yeu}, the primary objective of this paper is to constrain a physically reasonable $\eta^{(\prime)}$-meson LCWFs from their leading-twist LCDAs and, in conjunction with pQCD, to achieve a more precise determination of $\gamma\gamma^*\to\eta^{(\prime)}$ TFFs, e.g. $Q^2 F_{\eta\gamma}(Q^2)$ and $Q^2 F_{\eta'\gamma}(Q^2)$, up to next-to-leading order (NLO) QCD corrections. We will adopt the quark flavor mixing scheme to address specific theoretical challenges and update our earlier studies on $\gamma\gamma^*\to\eta^{(\prime)}$~\cite{Wu:2011gf}. Furthermore, by varying \(f_{\eta_{c_0}}\) within the frameworks of the $\eta$-$\eta'$-$\eta_c$ and $\eta$-$\eta'$-$G$-$\eta_c$ mixing mechanism, we explore how the charm quark and gluonic component influences the TFFs $Q^2 F_{\eta\gamma}(Q^2)$ and $Q^2 F_{\eta'\gamma}(Q^2)$.

The structure of this paper is organized as follows. In Section \ref{sec:2}, we provide a concise overview of the calculation techniques employed for the TFFs and introduce the LCWF models for $\eta_q$ and $\eta_s$ mesons -- defined within the quark flavor mixing scheme -- along with the analytical formalisms for their derivation. Section \ref{sec:3} presents our numerical results, where we analyze the properties of the TFFs $Q^2F_{\eta\gamma}(Q^2)$ and $Q^2F_{\eta'\gamma}(Q^2)$ in detail. Additionally, considering the possible intrinsic charm or gluonic components in the hadrons, we calculate the transformation matrices and their effects to the TFFs for both the $\eta$-$\eta'$-$\eta_c$ and $\eta$-$\eta'$-$G$-$\eta_c$ mixing mechanisms accordingly. Finally, Section \ref{sec:summary} provides a brief summary.

\section{Calculation Technology} \label{sec:2}

In this subsection, we begin with a mini-review of the key procedures for computing the TFFs for the $\gamma\gamma^*\to \eta^{(\prime)}$ process. In the quark-flavor basis, the two orthogonal basis states are assumed to possess the following parton-level composition within a Fock-state framework:
\begin{align}
|\eta_q\rangle=\frac{|u\bar u +d\bar d\rangle}{\sqrt{2}},~~~~~~|\eta_s\rangle=|s\bar s\rangle
\end{align}
The physical meson states $\eta$ and $\eta^\prime$ are related to the orthogonal states $\eta_q$ and $\eta_s$ through the following orthogonal transformation
\begin{align}
\begin{pmatrix} |\eta\rangle \cr |\eta'\rangle \end{pmatrix}
&= \begin{pmatrix} \cos\phi & -\sin\phi \cr
\sin\phi & \cos\phi \end{pmatrix} \begin{pmatrix} |\eta_q\rangle \cr |\eta_s\rangle \end{pmatrix},
\label{mixangle}
\end{align}
where $\phi$ is mixing angle. Under the quark-flavor mixing scheme, $\eta-\gamma$ and $\eta'-\gamma$ TFFs are related with $F_{\eta_q \gamma}(Q^2)$ and $F_{\eta_s \gamma}(Q^2)$ through the following equations
\begin{align}
F_{\eta\gamma}(Q^2)=F_{\eta_q \gamma}(Q^2)\cos\phi-F_{\eta_s \gamma}(Q^2)\sin\phi ,
\label{Eq:TFF1} \\
F_{\eta'\gamma}(Q^2)=F_{\eta_q \gamma}(Q^2)\sin\phi+F_{\eta_s \gamma}(Q^2)\cos\phi ,
\label{Eq:TFF2}
\end{align}
where $F_{\eta_q \gamma}(Q^2)$ and $F_{\eta_s \gamma}(Q^2)$ stand for $\eta_q \gamma$ and $\eta_s \gamma$ TFFs, respectively. Analogous to the pion-photon TFF, the meson-photon TFFs of $\eta_q$ and $\eta_s$ can also be decomposed into the following two parts
\begin{align}
F_{P\gamma}(Q^2)=F_{P\gamma}^{(V)}(Q^2)+F_{P\gamma}^{(NV)}(Q^2),
\end{align}
where for convenience, we have adopted the abbreviation $P$ ($P$ is a short notation for pseudoscalar) to stand for the $\eta_q$ and $\eta_s$-mesons. $F_{P\gamma}^{(V)}(Q^2)$ denotes the contribution from valence quarks, which involves the direct annihilation of a $q\bar{q}$ pair or $s\bar{s}$ pair into two photons; this constitutes the leading Fock-state contribution. $F_{P\gamma}^{(NV)}(Q^2)$ corresponds to the contribution from the non-valence quark components, which entails the coupling of one photon to the LCWF of $\eta_{q}$ and $\eta_{s}$-mesons -- specifically, this contribution arises from the strong-interaction-mediated photon couplings associated with the high-lying Fock-state components.

By keeping the transverse momentum dependence of the hard scattering amplitude and the LCWF, one can employ the pQCD approach in the light-cone framework to calculate the TFFs $F^{(V)}_{P\gamma}(Q^2)$~\cite{Lepage:1980fj}, which at present has been known up to the NLO level. Both the hard scattering amplitude and wavefunction contain for $k_\bot$-corrections, the TFFs for the valence quark state $F^{(V)}_{P\gamma}(Q^2)$ can be expressed as~\cite{Huang:2006wt}:
\begin{widetext}
\begin{equation}
F^{(V)}_{P\gamma}(Q^2) = 2\sqrt{3} e_P \int^1_0 dxdx'\delta(1-x-x') \times \int\frac{d^2{\mathbf k}_\bot}{16\pi^3}\Psi_P(x,{\mathbf k}_\bot)T_H(x,x',{\mathbf k}_\bot),
\end{equation}
\end{widetext}
where $e_P=(e^2_u+e^2_d,\sqrt{2}e^2_s)$ for $P=(\eta_q,\eta_s)$, respectively. The leading-order hard-scattering amplitude $T_H(x,x',{\mathbf k}_\bot)$ takes the form
\begin{align}
T_H(x,x',{\mathbf k}_\bot)=\frac{{\mathbf q}_\bot\cdot(x'{\mathbf q}_\bot+{\mathbf k}_\bot)}{{\mathbf q}^2_\bot(x'{\mathbf q}_\bot+{\mathbf k}_\bot)^2}+(x\leftrightarrow x')
\end{align}
with ${\mathbf k}_\bot$ being the transverse momentum of constituent quarks and ${\mathbf q}_\bot$ is transverse momentum of virtual photon. Regarding to the contribution from $F_{P\gamma}^{(NV)}(Q^2)$, its calculation proves challenging across any $Q^2$ region owing to its non-perturbative nature. To address this, one can devise a phenomenological model for $F_{P\gamma}^{(NV)}(Q^2)$ by relying on the asymptotic behavior observed as $Q^2\to 0$ and $Q^2\to\infty$. And the resulting expressions for the form factors $F^{(V)}_{P\gamma}(Q^2)$ and $F_{P\gamma}^{(NV)}(Q^2)$ can be written in the following form~\cite{Wu:2010zc, Huang:2006wt}:
\begin{widetext}
\begin{equation}
F_{P\gamma}^{(V)}(Q^2) = \frac{\sqrt{3}e_P}{4\pi^2} \int^1_0\int^{x^2Q^2}_0\frac{dx}{xQ^2} \left(1-\frac{C_F\alpha_s(Q^2)}{4\pi}\bigg(2 {\rm ln}x  +3+ {\rm ln} \frac{Q^2}{xQ^2+k^2_\bot}-\frac{\pi^2}{3}\bigg)\right)\Psi_P(x,k^2_\bot)dk^2_\bot
\label{TFF-BHL}
\end{equation}
\end{widetext}
and
\begin{equation}
F_{P\gamma}^{(NV)}(Q^2) = \frac{\alpha}{(1+Q^2/\kappa^2)^2},
\end{equation}
where the NLO QCD correction for the valence quark part is from Refs.\cite{Nandi:2007qx, Li:2009pr}. Here $C_F=4/3$ and $k_\bot=|{\mathbf k}_\bot|$. The aforementioned BHL prescription provides an explicit example of this assumption. Furthermore, $e_P$ is related to the electric charges of the constituent quarks, with the corresponding values being $e_{\eta_q}=5/9$ and $e_{\eta_s}=\sqrt{2}/9$. The parameters $\alpha$ and $\kappa$ are determined by
\begin{align}
&\alpha=\frac{1}{2}F_{P\gamma}(0), \\
&\kappa=\sqrt{-\frac{F_{P\gamma(0)}}{F_{P\gamma}^{(NV)'}|_{Q^2\to 0}}}.
\end{align}
The first order derivative for $F_{P\gamma}^{(NV)}(Q^2)$ with respect to $Q^2$ is given as the following expression
\begin{align}
&F_{P\gamma}^{(NV)'}|_{Q^2\to 0}
\nonumber\\
&=\frac{\sqrt{3}e_P}{8\pi^2}\bigg[\frac{\partial}{\partial Q^2}\int_0^1\int_0^{x^2Q^2}\bigg(\frac{\Psi_P(x,k^2_\bot)}{x^2Q^2}\bigg)dxdk^2_\bot\bigg]\bigg|_{Q^2\to 0}\label{Eq:dFP}
\end{align}
To compute the TFFs $F_{P\gamma}^{(V)}(Q^2)$ and $F_{P\gamma}^{(NV)}(Q^2)$ effectively, it is imperative to ascertain the precise light-cone wavefunction $\Psi_P(x,k^2_\bot)$. In our present discussions, without loss of generality, we have implicitly adopted the standard assumption that the LCWF depends on $\mathbf{k}_\perp$ only through $k_\perp^2$, i.e., $\Psi_{P}(x,\mathbf{k}_\perp) = \Psi_{P}(x,k_\perp^2)$~\cite{Huang:1994dy}.

As emphasized in Refs.\cite{Jakob:1994hd, Ong:1995gs, Cao:1996di, Li:2009pr}, it is crucial to retain the transverse-momentum dependence in both hard-scattering amplitude and the meson wavefunction to ensure a coherent and consistent analysis of the transition form factor. As for the pseudoscalar mesons $\eta_q$ and $\eta_s$, their LCWFs can be written as
\begin{align}
\Psi_P(x,{\mathbf k}_\bot)=\sum_{\lambda_1,\lambda_2}\chi^{\lambda_1,\lambda_2}(x,m_i,{\mathbf k}_\bot)\Psi_P^R(x,m_i,{\mathbf k}_\bot).
\end{align}
Here $m_i$ represents light quark $m_q$ or $m_s$, $\lambda_1$ and $\lambda_2$ denote the helicities of two constituent quarks. The spin-space wavefunction $\chi^{\lambda_1,\lambda_2}(x,{\mathbf k}_\bot)$ is derived from Wigner-Melosh rotation. Since the contribution of higher helical state $\lambda_1+\lambda_2=\pm 1$ is little compared to that of the usual helical state $\lambda_1+\lambda_2=0$, we only write explicit terms for the usual helical state, {\it i.e.} $\chi^{\lambda_1,\lambda_2}(x,m_i,{\mathbf k}_\bot)=m_i/ \sqrt{2(m^2_i+k^2_\bot)}$~\cite{Wu:2010zc, Wu:2011gf, Huang:2004su}. Following the BHL prescription, we can express the spatial wavefunction $\Psi_P^R(x,m_i,{\mathbf k}_\bot)$ as~\cite{Huang:1994dy, Cao:1997hw}
\begin{align}
\Psi_P^R(x,m_i,{\mathbf k}_\bot)=A_P\varphi_P(x)\exp\bigg[-\frac{k^2_\bot+m_i^2}{8\beta_P^2x \bar x }\bigg],
\end{align}
where $\bar x = (1-x)$. $A_P$ denotes the overall normalization parameter. The $x$-dependence function $\varphi_P(x)$, which dominates the longitudinal behavior of the wavefunction, remains to be determined. Here, we adopt the light-cone harmonic oscillator (LCHO) model to parameterize this behavior. And following the discussions in Ref.~\cite{Zhong:2021epq}, we take its form as follows:
\begin{align}
\varphi_P(x)&=[x\bar x]^{\epsilon_P}(1+\sigma_P\times C_2^{3/2}(\xi)),
\end{align}
with $\xi = (2x-1)$. The $C^{2/3}_n$ stand for $n$th-order Gegenbauer polynomial~\footnote{In addition to the present adopted model for $\varphi_P(x)$, other models have also been employed in the literature. It has been shown in Refs.\cite{Zhong:2021epq, Hu:2024tmc}, if the pseudoscalar LCDAs exhibit a similar unimodal profile, different LCDA model choices will yield numerically close results for the resulting TFFs.}. Then, the specific expression for the light-cone wavefunction is:
\begin{align}
\Psi_P(x,{\mathbf k}_\bot)&=\frac{m_i A_P}{\sqrt{2(m^2_i+k^2_\bot)}}[x\bar x]^{\epsilon_P} (1+\sigma_P\times C_2^{3/2}(\xi)) \nonumber\\
&\times\exp\bigg[-\frac{k^2_\bot+m_i^2}{8\beta_P^2x \bar x }\bigg]
\end{align}
To fix the expression for the LCWF, we need to determine the values of parameters $A_P, \beta_P$, and $\epsilon_P, \sigma_P$.

There is a relationship between the leading-twist LCDA $\phi_{2;P}(x,\mu)$ and the LCWF $\Psi_P(x,{\mathbf k}_\bot)$, {\it i.e.}
\begin{align}
\phi_{2;P}(x,\mu)=\frac{2\sqrt{6}}{f_P}\int_{|{\mathbf k}_\bot|^2<\mu^2}\frac{d^2{\mathbf k}_\bot}{16\pi^3}\Psi_P(x,{\mathbf k}_\bot).
\end{align}
Upon integrating over the transverse momentum $\mathbf{k}_\perp$, the expression for the leading-twist LCDA $\phi_{2;P}(x,\mu)$ of $\eta_{q,s}$ mesons is then given by
\begin{align}
&\phi_{2;P}(x,\mu) =\frac{\sqrt{3} A_P m_i\beta_P}{2\sqrt {2}\pi^{3/2}f_P}\sqrt{x\bar x} \varphi_P(x)
\nonumber\\
&\hspace{1.3cm}\times\bigg({\rm Erf}\bigg[\sqrt{\frac{m_i^2+\mu^2}{8\beta^2_Px\bar x}}\bigg]-{\rm Erf}\bigg[\sqrt{\frac{m_i^2}{8\beta^2_P x\bar x}}\bigg]\bigg)
\label{Eq:DA2}
\end{align}
where the error function ${\rm Erf}(x)$ is defined as ${\rm Erf}(x)=\frac{2}{\sqrt{\pi}}\int_0^x e^{-t^2}dt$. The moments
\begin{displaymath}
\langle\xi^P_n\rangle|_{\mu}=\int_{0}^{1}(2x-1)^{n} \phi_{2;P}(x,\mu)
\end{displaymath}
can be derived via the QCD sum rule approach within the background field theory~\cite{Hu:2023pdl, Zhang:2025yeu}. The LCWF should satisfy the following two constraints:
\begin{itemize}
\item The wavefunction normalization condition
\begin{align}
\int_0^1 dx\int \frac{d^2{\mathbf k}_\bot}{16\pi^3}\Psi_P(x,{\mathbf k}_\bot)=\frac{f_P}{2\sqrt{6}}.
\label{Eq:NC}
\end{align}
\item The sum rule derived from the decay of pseudoscalar mesons into two photons~\cite{bhl}
\begin{align}
\int_0^1 dx\Psi_P(x,{\mathbf k}_\bot=0)=\frac{\sqrt{6}}{f_P}.
\label{Eq:da}
\end{align}
\end{itemize}

\section{Numerical Analysis}\label{sec:3}

We adopt the NNLO $\alpha_s(Q^2)$ to do the numerical calculation, {\it i.e.}
\begin{eqnarray}
\alpha_s(Q^2) &=& \frac{1}{\beta_0 \ln(Q^2/{\Lambda^2_{\rm QCD}})}\bigg\{1-\frac{\beta_1}{\beta^2_0}\frac{\ln[\ln(Q^2/{\Lambda^2_{\rm QCD}})]}{\ln(Q^2/{\Lambda^2_{\rm QCD}})}
\nonumber\\
& & +\frac{1}{\beta^4_0 \ln^2(Q^2/{\Lambda^2_{\rm QCD}})}\bigg\{ \beta^2_1\{\ln^2[\ln(Q^2/{\Lambda^2_{\rm QCD}})]
\nonumber\\
& & -\ln[\ln(Q^2/{\Lambda^2_{\rm QCD}})]-1\} +\beta_0\beta_2\bigg\}\bigg\},
\end{eqnarray}
where
\begin{align}
\beta_0&=\frac{33-2 n_f}{12\pi},  \nonumber\\
\beta_1&=\frac{153-19 n_f}{24\pi^2},  \nonumber\\
\beta_2&=\frac{2857-5033 n_f/9+325 n^2_f/27}{128\pi^3},
\end{align}
where $n_f$ represents the active number of light quarks, and its magnitude varies with energy scale. $\Lambda_{\rm QCD}$ denotes the asymptotic scale of QCD and its value can be determined by requiring $\alpha_s(m_Z)=0.118$~\cite{ParticleDataGroup:2024cfk}. The mixing angle $\phi$ and the $\eta_q, \eta_s$ meson decay constants $f_{\eta_q}, f_{\eta_s}$ have been calculated under the QCD sum rule approach, namely $\phi={41.2^\circ}^{+0.05^\circ}_{-0.06^\circ}$~\cite{Hu:2021zmy}, $f_{\eta_q}=0.141~{\rm GeV}$~\cite{Hu:2023pdl} and $f_{\eta_s}=0.176~{\rm GeV}$~\cite{Zhang:2025yeu}. As for the moments $\langle\xi^P_n\rangle$ (where $P=\eta_q$ or $\eta_s$), their first three non-zero terms are given in our recent works~\cite{Hu:2023pdl, Zhang:2025yeu}:
\begin{align}
\langle\xi^{\eta_q}_2\rangle|_{\mu_0}=0.253\pm0.016, ~~\langle\xi^{\eta_s}_2\rangle|_{\mu_0}=0.194\pm0.008,
\nonumber\\
\langle\xi^{\eta_q}_4\rangle|_{\mu_0}=0.127\pm0.010, ~~\langle\xi^{\eta_s}_4\rangle|_{\mu_0}=0.087\pm0.006,
\nonumber\\
\langle\xi^{\eta_q}_6\rangle|_{\mu_0}=0.080\pm0.008, ~~\langle\xi^{\eta_s}_6\rangle|_{\mu_0}=0.051\pm0.007, \nonumber
\end{align}
where the initial scale $\mu_0=1~{\rm GeV}$.

\subsection{Basic results}

\begin{table}[htb]
\renewcommand\arraystretch{1.3}
\center
\footnotesize
\caption{The parameters $A_P$ (in unit: ${\rm GeV^{-1}}$), $\beta_P$ (in unit: ${\rm GeV}$), $\epsilon_P$ and $\sigma_P$ that are derived under several typical choices of constituent quark masses $m_q$ and $m_s$ (in unit: ${\rm MeV}$), respectively.}\label{Tab:LCHO}
\begin{tabular}{lllll }
\hline
~~$m_q$~~   ~~~~& $A_{\eta_q}$~~~~   ~~~~& $\beta _{\eta_q}$~~~~
~~~~& $\epsilon_{\eta_q}$~~~~    ~~~~& $\sigma_{\eta_q}$~~~~      \\   \hline
~~$250$    & $9.461$     & $0.895$   & $-0.631$   & $-0.116$          \\
~~$300$    & $6.979$     & $0.759$   & $-0.836$   & $-0.114$         \\
~~$350$    & $5.260$     & $0.684$   & $-1.044$   & $-0.112$          \\   \hline
~~$m_s$~~~   ~~~~& $A_{\eta_s}$~~~~   ~~~~& $\beta _{\eta_s}$~~~~
~~~~& $\epsilon_{\eta_s}$~~~~    ~~~~& $\sigma_{\eta_s}$~~~~      \\   \hline
~~$400$    & $22.699$   & $1.010$   & $-0.065$   & $-0.108$          \\
~~$450$    & $18.552$    & $0.899$   & $-0.225$   & $-0.106$         \\
~~$500$    & $15.256$     & $0.829$   & $-0.394$   & $-0.105$          \\
\hline
\end{tabular}
\end{table}

\begin{figure}[htb]
\begin{center}
\includegraphics[width=0.45\textwidth]{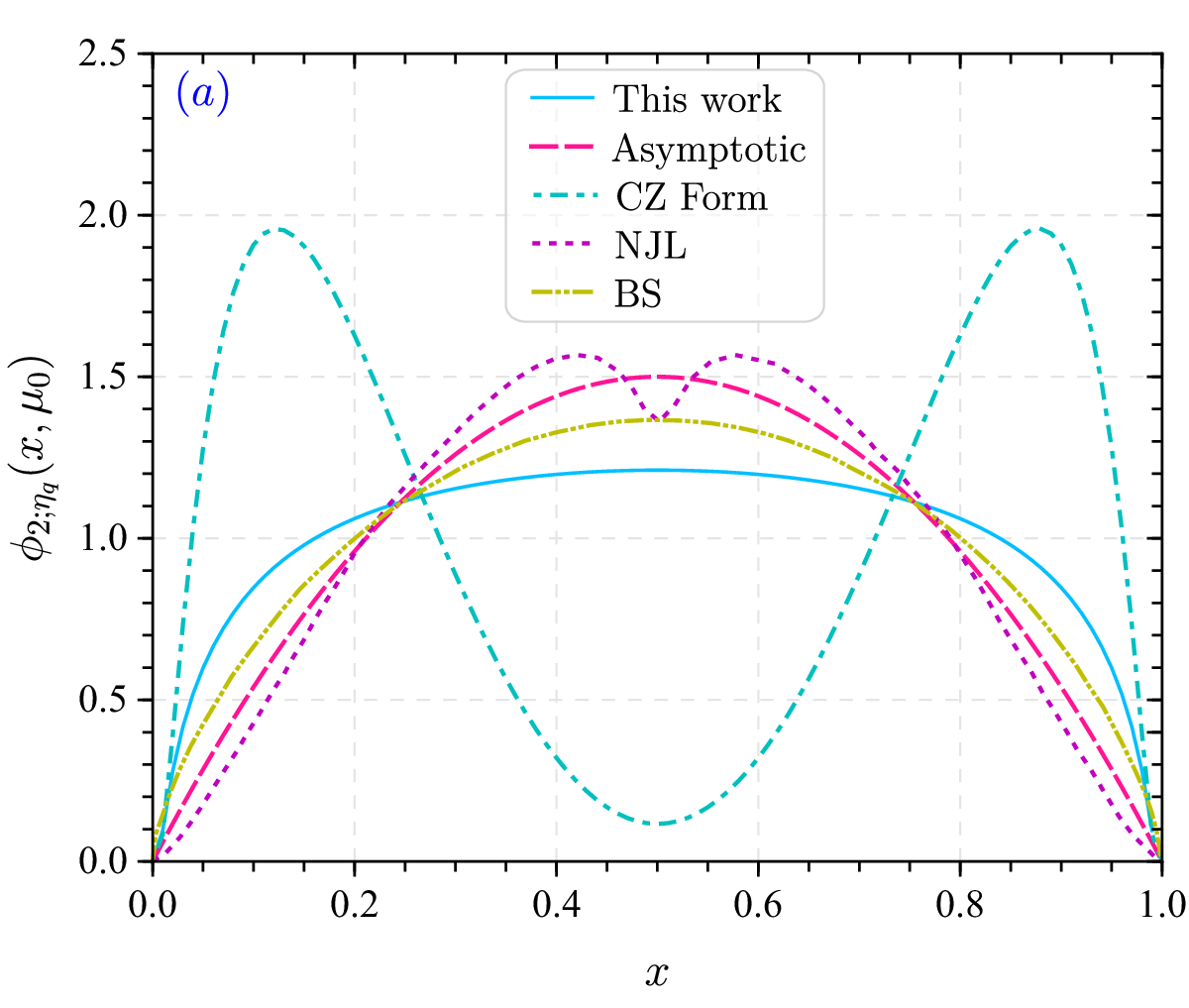}
\includegraphics[width=0.45\textwidth]{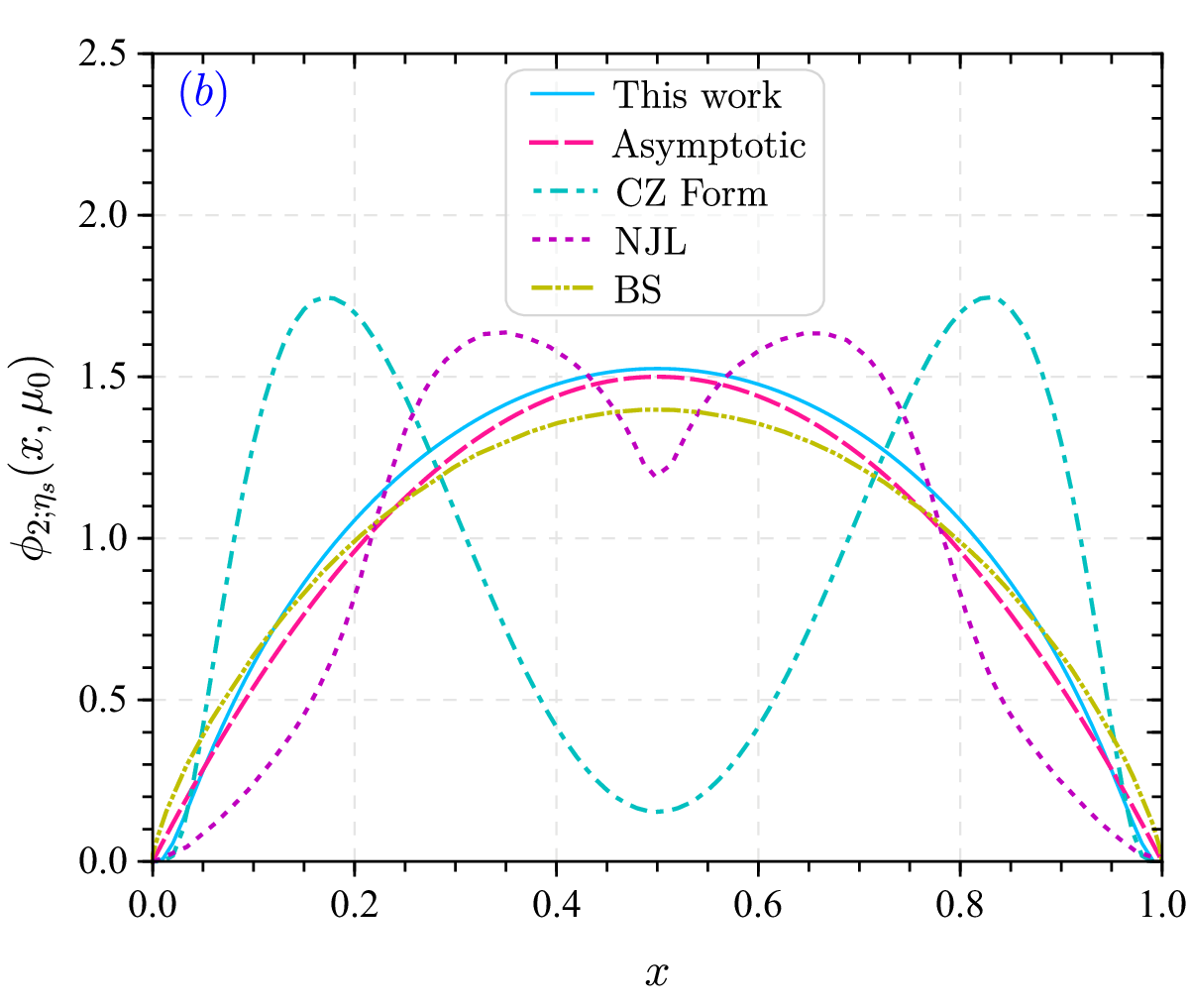}
\end{center}
\caption{The $\eta_q$ and $\eta_s$ mesons' leading-twist LCDAs $\phi_{2;\eta_q}(x,\mu_0) (a)$ and $\phi_{2;\eta_s}(x,\mu_0) (b)$ with $\mu_0=1~{\rm GeV}$. As a comparison, the curves from the nonlocal NJL model~\cite{GomezDumm:2016bxp}, the BS model~\cite{Ding:2018xwy}, the CZ-form and the asymptotic form have also been presented, respectively.} \label{fig:DA}
\end{figure}

Using the above-provided $\langle\xi^P_n\rangle$-moments, together with the constraints (\ref{Eq:NC}, \ref{Eq:da}), we first fix all the required input parameters for the LCWF, such as $\epsilon_P$, $\sigma_P$, $A_P$, and the harmonic parameter $\beta_P$, via the method of least squares~\cite{Zhong:2021epq}. The results are presented in Table~\ref{Tab:LCHO}, where results for several typical choices of constituent quark masses are shown. The LCDAs $\phi_{2;\eta_q}(x, \mu_0)$ and $\phi_{2;\eta_s}(x, \mu_0)$ can be fixed accordingly; the LCDAs with all input parameters set to their central values are shown in Fig.~\ref{fig:DA}. For comparison, we also present the LCDAs from the nonlocal Nambu–Jona-Lasinio (NJL) model~\cite{GomezDumm:2016bxp}, the Bethe-Salpeter (BS) model~\cite{Ding:2018xwy}, the CZ-form, and the asymptotic form in Fig.~\ref{fig:DA}, respectively. The NJL LCDAs exhibit a CZ-like double-peaked behavior, which arises from contributions generated by the nonlocal character of the interactions; And the LCDAs of Ref.\cite{Ding:2018xwy}, based on a BS kernel contribution model of non-Abelian anomalies, favor an asymptotic-like single-peaked behavior. Our present LCHO model is also asymptotic-like, and due to the extra exponential suppression, as shown in Eq.(\ref{Eq:DA2}), it is slightly sharper than the asymptotic LCDA at the endpoints $x\to 0$ or $x\to 1$.

\begin{figure}[htb]
\begin{center}
\includegraphics[width=0.45\textwidth]{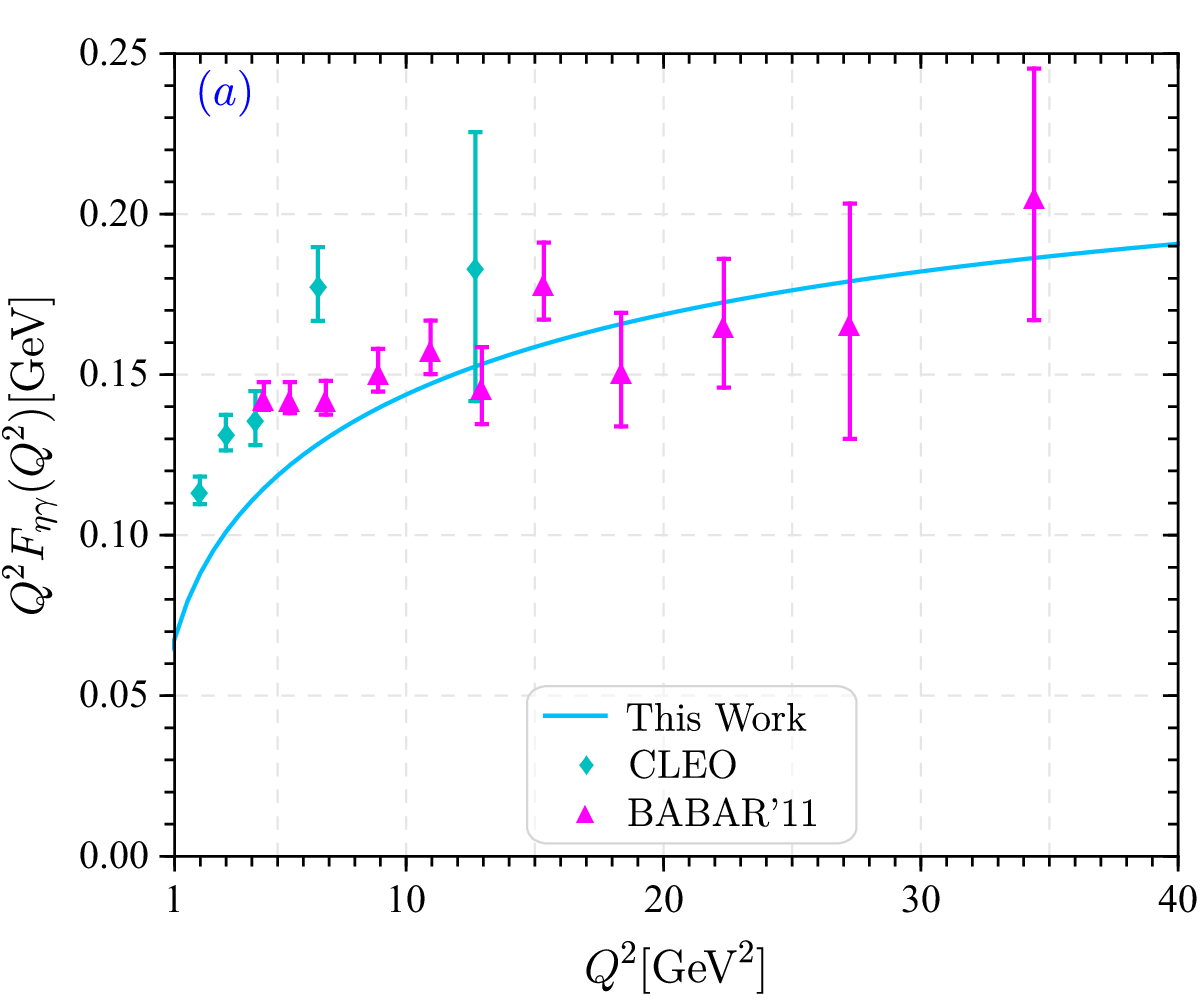}
\includegraphics[width=0.45\textwidth]{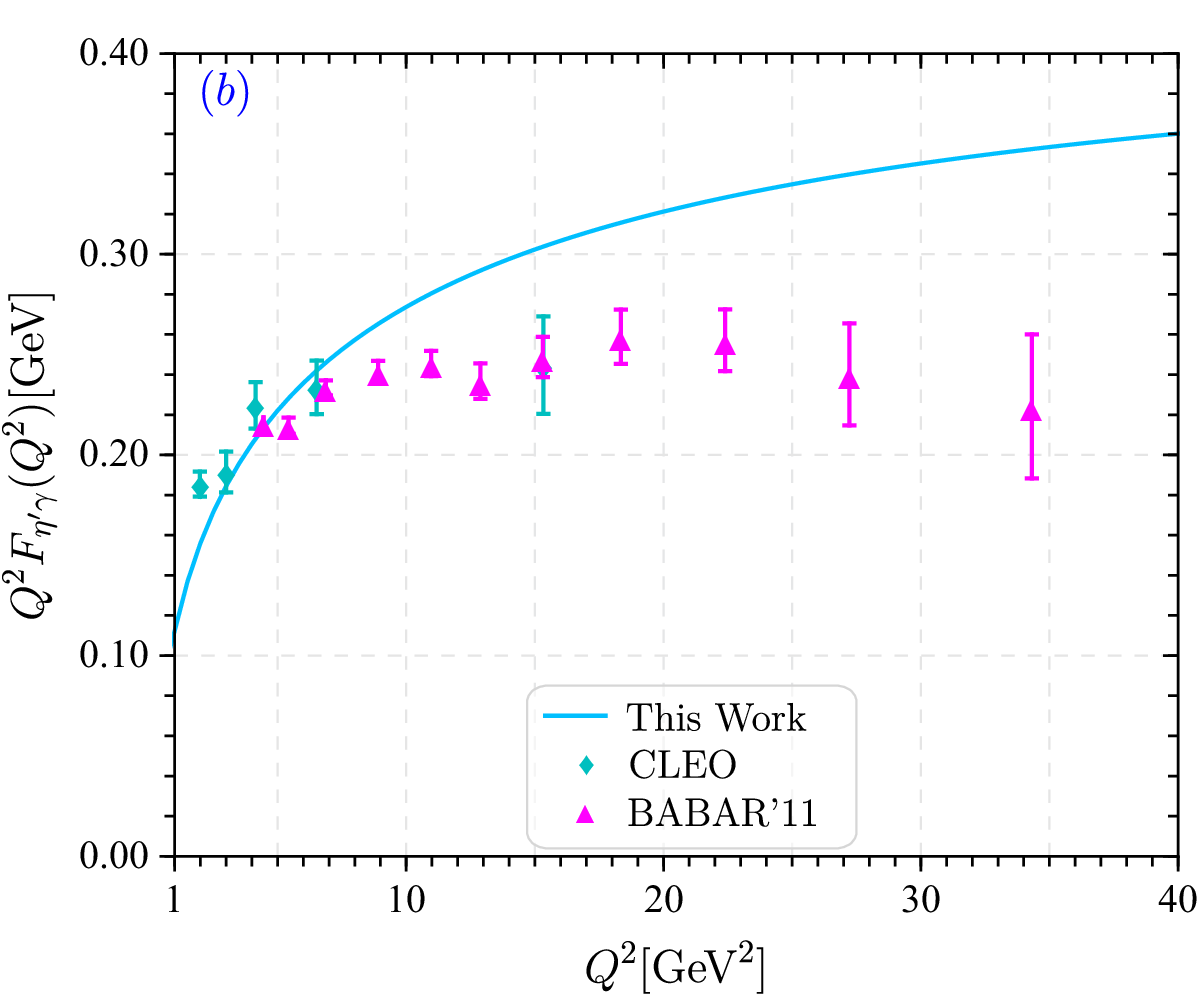}
\end{center}
\caption{The $\eta-\gamma$ and $\eta'-\gamma$ TFFs $Q^2 F_{\eta\gamma}(Q^2)$ (a) and $Q^2 F_{\eta'\gamma}(Q^2)$ (b). For comparison, the experimental results from CLEO~\cite{CLEO:1997fho} and BABAR'11~\cite{BaBar:2011nrp} are also represented.}
\label{fig:Tff}
\end{figure}

Using the formulas (\ref{Eq:TFF1}, \ref{Eq:TFF2}) for the TFFs under the standard quark-flavor mixing scheme and the LCHO model (\ref{Eq:DA2}) for the $\eta^{(\prime)}$-LCDAs, we calculate the TFFs $Q^2 F_{\eta\gamma}(Q^2)$ and $Q^2 F_{\eta'\gamma}(Q^2)$ and present them in Fig.~\ref{fig:Tff}. These curves are calculated by fixing all input parameters to be their central values -- specifically $\phi=41.2^\circ$, $m_q=0.30~{\rm GeV}$ and $m_s=0.45~{\rm GeV}$. The experimental data of CLEO~\cite{CLEO:1997fho} and BABAR'11~\cite{BaBar:2011nrp} are also presented. Fig.~\ref{fig:Tff} shows that the predicted $Q^2 F_{\eta\gamma}(Q^2)$ is consistent with the BABAR'11 data in both intermediate and large $Q^2$ regions, and the predicted $Q^2 F_{\eta'\gamma}(Q^2)$ is consistent with the CLEO data in low $Q^2$ region. But in other regions, significant discrepancies are observed. Thus, it is crucial to demonstrate whether these two TFFs can be accounted for concurrently through appropriate selection of the input parameters.

\begin{figure}[htb!]
\begin{center}
\includegraphics[width=0.45\textwidth]{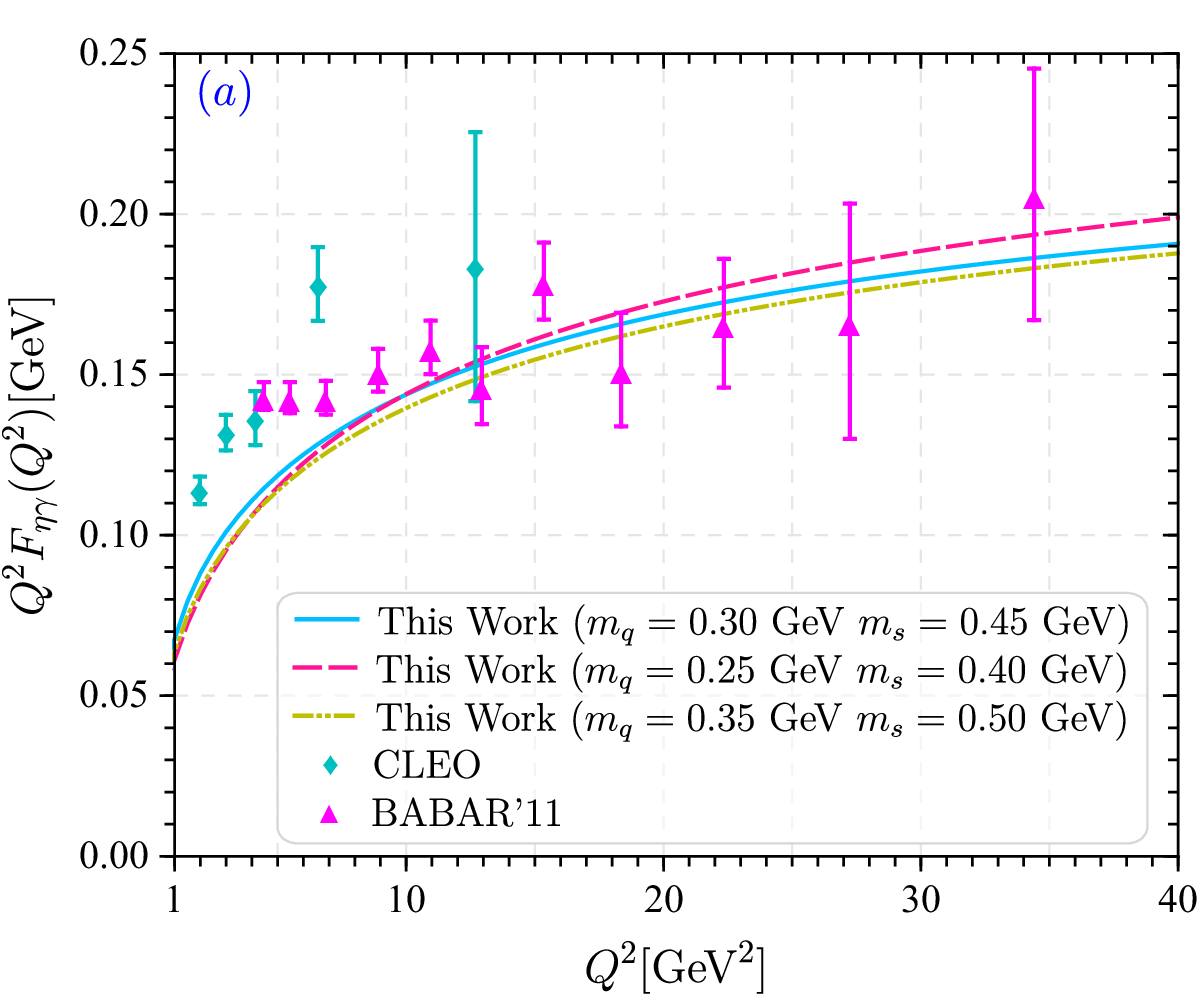}
\includegraphics[width=0.45\textwidth]{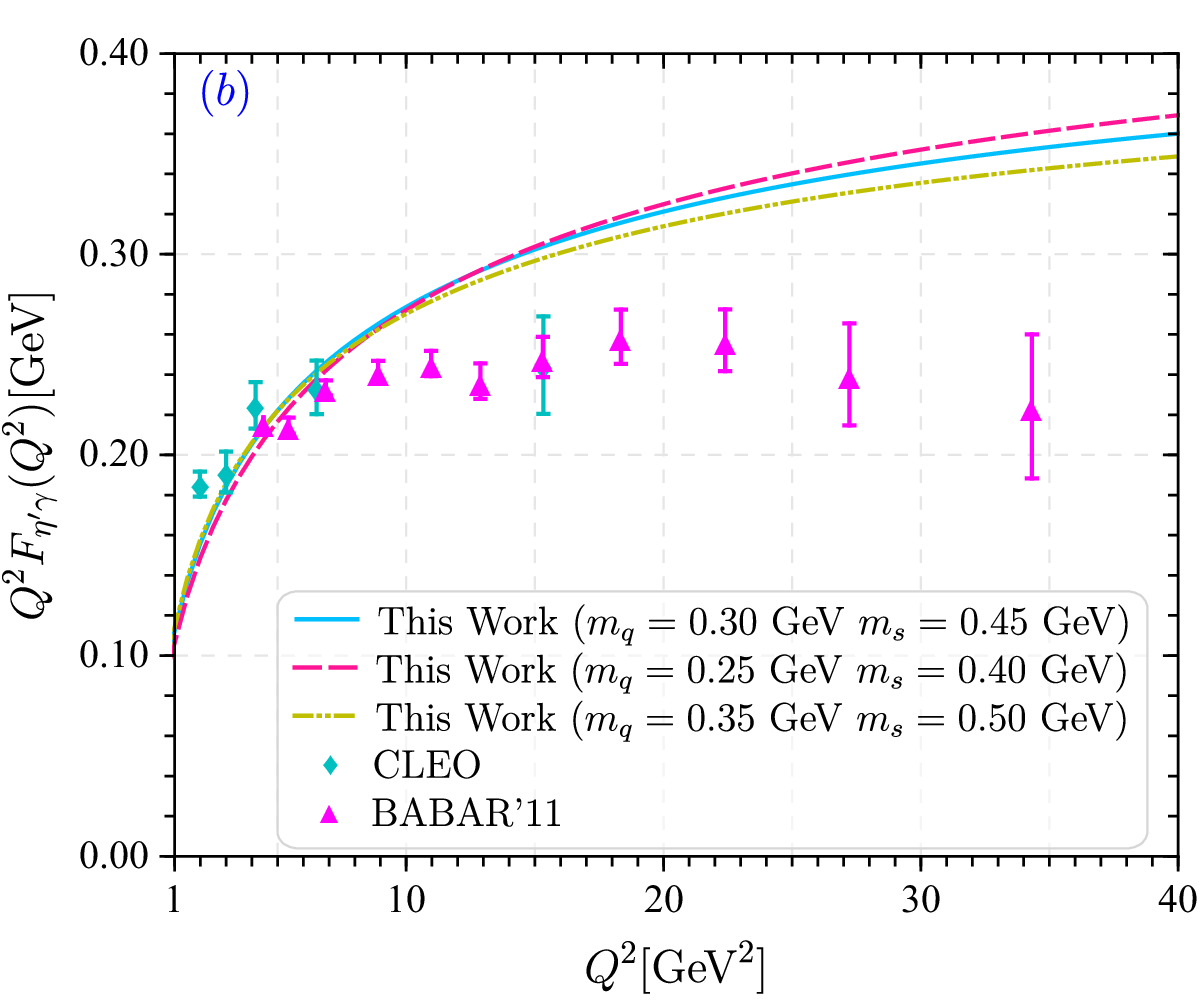}
\end{center}
\caption{The $\eta-\gamma$ and $\eta'-\gamma$ TFFs $Q^2F_{\eta\gamma}(Q^2)$ (a) and $Q^2F_{\eta'\gamma}(Q^2)$ (b) are calculated for different mixing angles. The experimental results of CLEO~\cite{CLEO:1997fho} and BABAR'11~\cite{BaBar:2011nrp} are also represented.} \label{fig:mqs}
\end{figure}

Subsequently, we delve into the uncertainties in the TFFs arising from variations in the mixing angle $\phi$, $m_q$, and $m_s$, which are the main error sources for these TFFs. We first consider the uncertainties from different choices of $\phi$. Considering that the error in the mixing angle $\phi$ is small~\cite{Hu:2021zmy}, its impact on the numerical results of the TFFs is only at the thousandths place, which has a negligible effect on the final outcome. We then consider the uncertainties caused by taking $m_q=0.30\pm0.05~{\rm GeV}$ and $m_s=0.45\pm0.05~{\rm GeV}$, while keeping the mixing angle $\phi$ fixed at $\phi=41.2^\circ$. The $\eta^{(\prime)}-\gamma$ TFFs under three cases, e.g. $[m_q=0.25~{\rm GeV}$ and $m_s=0.40~{\rm GeV}]$, $[m_q=0.30~{\rm GeV}$ and $m_s=0.45~{\rm GeV}]$ and $[m_q=0.35~{\rm GeV}$ and $m_s=0.50~{\rm GeV}]$, are presented in Fig.~\ref{fig:mqs}. Fig.~\ref{fig:mqs} shows the TFFs demonstrate a dependence on the quark mass variations. Specifically, in higher $Q^2$ region, both TFFs exhibit a downward trend with the increment of constituent quark masses.

\subsection{An improved analysis by taking intrinsic charm and gluonic components into consideration}

The above results indicate that employing the $\eta_q$-$\eta_s$ mixing scheme alone makes it difficult to consistently explain both TFFs $Q^2 F_{\eta\gamma}(Q^2)$ and $Q^2 F_{\eta'\gamma}(Q^2)$ concurrently. To narrow the gap between the predicted TFFs and experimental data, we further incorporate the contributions from intrinsic charm and gluonic components as a next step.

In the physical states $|\eta\rangle$ and $|\eta'\rangle$, there may exist intrinsic charm and the gluonic components. Then one can extend the matrix (\ref{mixangle}) to include the physical pseudoscalar glueball $|G\rangle$ and pseudoscalar meson $|\eta_c\rangle$ components. For the possible two types of mixing, e.g. $\eta-\eta'-\eta_c$ and $\eta-\eta'-G-\eta_c$, the physical states $|\eta\rangle$, $|\eta'\rangle$, $|G\rangle$ and $|\eta_c\rangle$ are related to the unmixed $|\eta_q\rangle$, $|\eta_s\rangle$, glueball state $|gg\rangle$ and intrinsic charm component $|\eta_{c_0}\rangle$ via the following rotations~\cite{Zhang:2025yeu}
\begin{align}\label{eq:qsc}
	\left( \begin{array}{c}
		|\eta\rangle \\ |\eta^\prime\rangle\\|\eta_c\rangle
	\end{array} \right)
	= U(\theta,\theta_c,\theta_i)
	\left( \begin{array}{c}
		|\eta_q\rangle \\ |\eta_s\rangle\\|\eta_{c_0}\rangle
	\end{array} \right),
\end{align}
\begin{align}\label{eq:qsgc}
	\left( \begin{array}{c}
		|\eta\rangle \\ |\eta^\prime\rangle\\|G\rangle\\|\eta_c\rangle
	\end{array} \right)
	=U(\theta,\phi_g,\phi_c,\theta_g)
	\left( \begin{array}{c}
		|\eta_q\rangle \\ |\eta_s\rangle\\|gg\rangle\\|\eta_{c_0}\rangle
	\end{array} \right),
\end{align}
where
\begin{eqnarray}
U(\theta,\theta_c,\theta_i) &=& U_{3}(\theta)U_{1}(\theta_c)U_{3}(\theta_i), \nonumber\\
U(\theta,\phi_g,\phi_c,\theta_g) &=& U_{34}(\theta)U_{14}(\phi_g)U_{13}(\phi_c)U_{12}(\theta_g)U_{34}(\theta_i) \nonumber .
\end{eqnarray}
Here $\theta$, $\theta_c$, $\phi_g$, $\phi_c$ and $\theta_g$ are mixing angles among different components; and the ideal mixing angle $\theta_i=54.7^\circ$~\cite{Cao:2012nj}. Substituting the transformation matrices $U_{1,3}$, $U_{12}$, $U_{13}$, $U_{14}$ and $U_{34}$ into the above equations, whose explicit forms are presented in Appendix A, we obtain
\begin{widetext}
\begin{align}
	\setlength{\arraycolsep}{1.2pt}
	U(\theta,\theta_c,\theta_i) =
	\begin{pmatrix}
		c\theta c\theta_i-s\theta s\theta_i c\theta_c &~~ -s\theta c\theta_i c\theta_c-c\theta s\theta_i &~~ -s\theta s\theta_c \\
		c\theta s\theta_i c\theta_c + s\theta c\theta_i &~~ c \theta c\theta_i c\theta_c-s\theta s\theta_i &~~ c\theta s\theta_c  \\
		-s\theta_i s\theta_c & -c\theta_i s\theta_c & c\theta_c
	\end{pmatrix}
	\label{eq:uqsc}
\end{align}
\begin{align}
	\setlength{\arraycolsep}{1.2pt}
	U(\theta,\phi_g,\phi_c,\theta_g) 	&=
	\begin{pmatrix}
		c\theta c\theta_i-s\theta s\theta_i c\phi_c c\phi_g & -s\theta c\theta_i c\phi_c c\phi_g-c\theta s\theta_i & s\theta\left(s\theta_g s\phi_c c\phi_g-c\theta_g s\phi_g\right) & -s\theta \left(c\theta_g s\phi_c c\phi_g +s\theta_g s\phi_g \right) \\
		c\theta s\theta_i c\phi_c c\phi_g + s\theta c\theta_i & c \theta c\theta_i c\phi_c c\phi_g-s\theta s\theta_i & c\theta\left( c\theta_g s\phi_g - s\theta_g s\phi_c \right) c\phi_g & c\theta \left( c\theta_g s\phi_c c\phi_g + s\theta_g s\phi/_g\right)  \\
		-s\theta_i c\phi_c s\phi_g & -c\theta_i c\phi_c s\phi_g & s\theta_g s\phi_c s\phi_g+c\theta_g c\phi_g  & s\theta_g c\phi_g -c\theta_g s\phi_c s\phi_g  \\
		-s\theta_i s\phi_c & -c\theta_i s\phi_c & -s\theta_g c\phi_c & c\theta_g c\phi_c
	\end{pmatrix}
	\label{eq:uqsgc}
\end{align}
\end{widetext}
where for convenience, we have adopted the short notations, $c \equiv \cos(\cdot)$ and $s \equiv \sin(\cdot)$. Together with the constraints on the related mass matrix elements, these mixing angles can be numerically fixed in a recursive, step-by-step manner. To shorten the length, we only list the results here, and interested readers may refer to Ref.\cite{Zhang:2025yeu} for details.

As for the $\eta -\eta^\prime-\eta_c$ mixing scheme, the required mixing angles are $\theta=(-14.268)^{\circ}$ and $\theta_c=(-1.025)^{\circ}$, which lead to
\begin{align}	
	U(\theta,\theta_c,\theta_i)=
	\begin{pmatrix}
		0.7611 & -0.6486 & -0.0044  \\
		0.6484 & 0.7611  & -0.0173   \\
		0.0146 & 0.0103  & 0.9998
	\end{pmatrix}.
\label{eq:fqsc}
\end{align}
As for the $\eta-\eta^\prime-G-\eta_c$ mixing scheme, the required mixing angles are $\theta=(-13.441)^{\circ}$, $\theta_g=(-0.005)^{\circ}$, $\phi_c=(-1.213)^{\circ}$ and $\phi_g=(-0.133)^{\circ}$, which lead to
\begin{align}	
	U(\theta,\phi_g,\phi_c,\theta_g)=
	\begin{pmatrix}
		0.7517 & -0.6595 & -0.0005 & -0.0049 \\
		0.6593 & 0.7516 & -0.0023 & -0.0206  \\
		0.0019 & 0.0013 & 1.000  & -0.0001 \\
		0.0173 & 0.0122 & 0.0001 & 0.9998
	\end{pmatrix}.
\label{eq:fqsgc}
\end{align}

Next, we improve our treatment of the TFFs within these two mixing schemes. For the $\eta$-$\eta'$-$\eta_c$ mixing scheme, the corresponding TFFs can be derived via the transformation matrix \eqref{eq:fqsc}, i.e.
\begin{eqnarray}
	F_{\eta\gamma}(Q^2) &=& 0.7611\times F_{\eta_q \gamma}(Q^2)-0.6486\times F_{\eta_s \gamma}(Q^2)
	\nonumber\\
	& & -0.0044\times F_{\eta_{c_0}\gamma}(Q^2), \label{Eq:eta11} \\
	F_{\eta'\gamma}(Q^2) &=& 0.6484\times F_{\eta_q \gamma}(Q^2)+0.7611\times F_{\eta_s \gamma}(Q^2)
	\nonumber\\
	& & -0.0173\times F_{\eta_{c_0}\gamma}(Q^2),
	\label{Eq:eta12}
\end{eqnarray}
\begin{figure*}[htb]
	\begin{center}
		\includegraphics[width=0.45\textwidth]{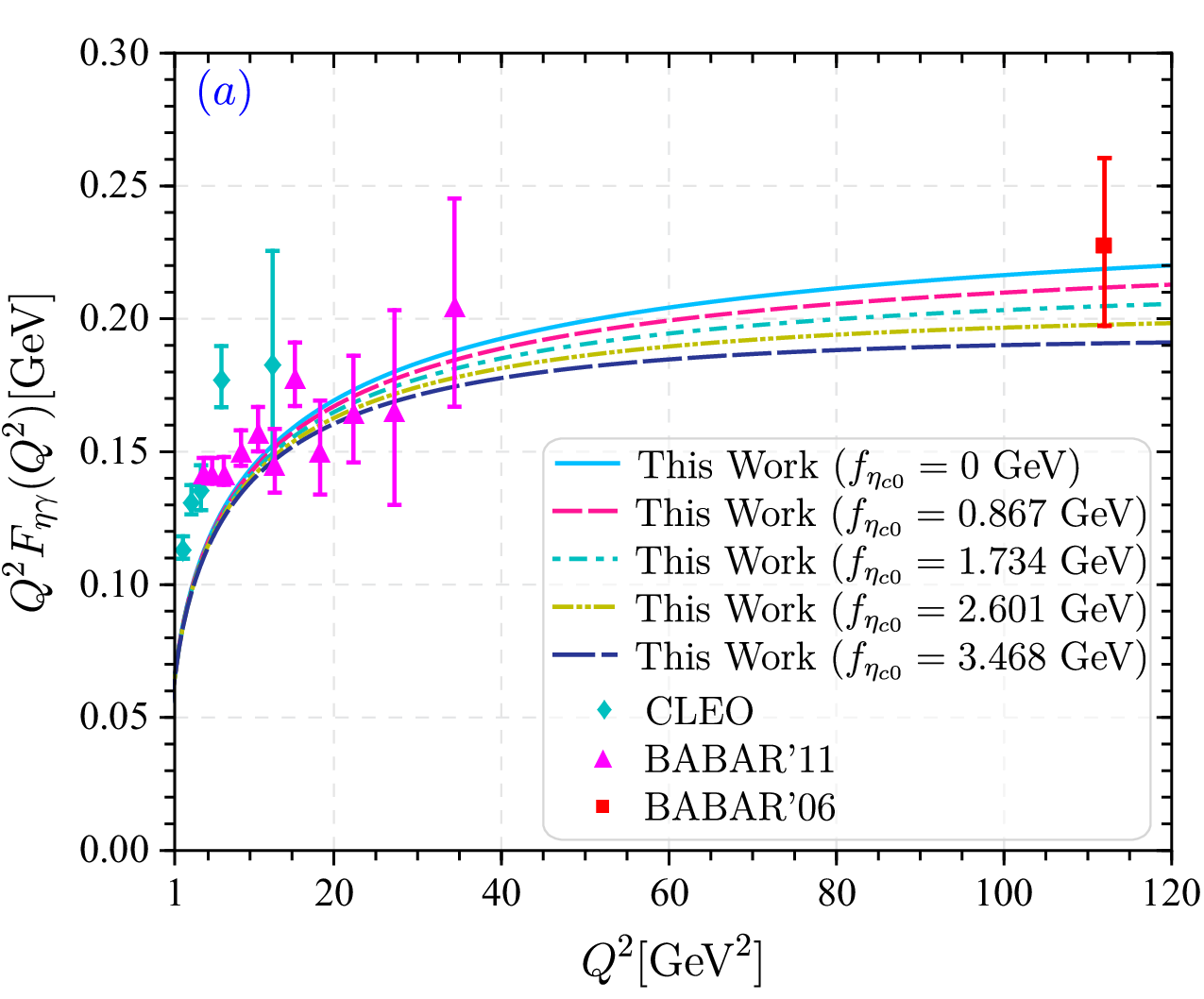}
		\includegraphics[width=0.45\textwidth]{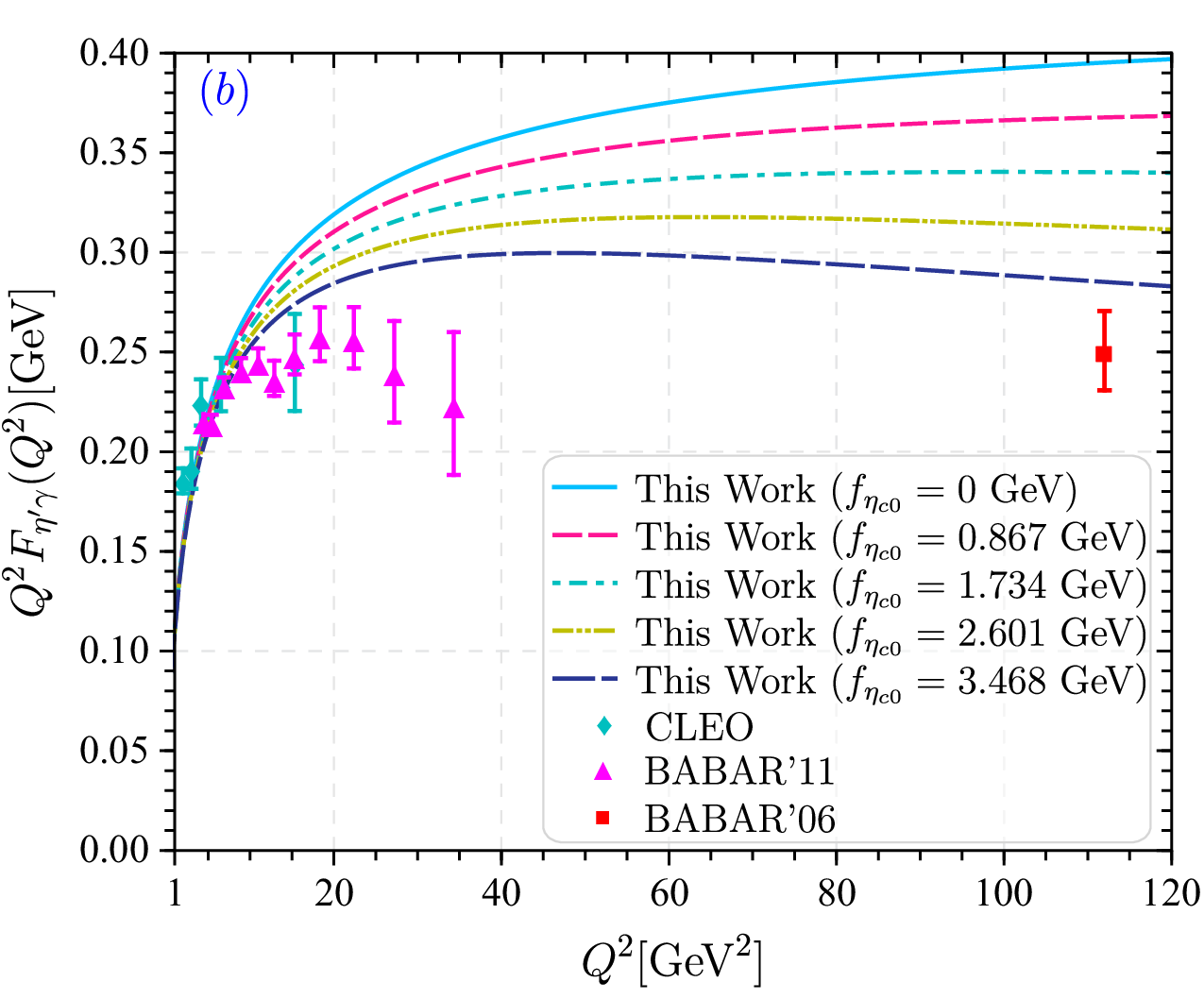}
	\end{center}
	\caption{The $\eta-\gamma$ and $\eta'-\gamma$ TFFs $Q^2F_{\eta\gamma}(Q^2)$ (a) and $Q^2F_{\eta'\gamma}(Q^2)$ (b) are obtained through $\eta-\eta'-\eta_c$ mixing mechanism. The experimental results from CLEO~\cite{CLEO:1997fho}, BABAR'11~\cite{BaBar:2011nrp} and BABAR'06~\cite{BaBar:2006ash} are also represented.} \label{fig:fetaetac}
\end{figure*}
where $F_{\eta_{c_0}\gamma}(Q^2)$ represents the contribution from unmixed intrinsic charm component in $\eta$ and $\eta^\prime$.
Here, the calculation of $F_{\eta_{c_0}\gamma}(Q^2)$ requires the inclusion of the charm quark mass effect in the hard part of the amplitude. Specifically, the higher-helicity states, which are proportional to the constituent quark mass, contribute significantly. Upon integrating over the azimuth angle, a direct calculation shows~\cite{Huang:2006as}
\begin{align}
Q^2 F_{\eta_{c_0}\gamma}(Q^2)&=\frac{\sqrt{2}}{3\sqrt{3}\pi^2} \int^1_0\frac{dx}{x}\int^{\infty}_0 \Psi_{\eta_{c_0}}({x,{\mathbf k}_\bot})\bigg[1+
\nonumber\\
&~~\frac{1-z-y^2}{\sqrt{(z+(1-y)^2)(z+(1+y)^2)}}k_\bot d k_\bot\bigg]
\label{Eq:etac}
\end{align}
where $z=\frac{m_c^2}{x^2Q^2}$ and $y=\frac{k_\bot}{xQ}$. To maintain consistency, we also adopt the LCHO model for the light-cone wavefunction of the $\eta_{c_0}$ meson, e.g.
\begin{align}
\Psi_{\eta_{c_0}}({x,{\mathbf k}_\bot})& = A_{\eta_{c_0}}\left([x\bar x]^{\epsilon_{\eta_{c_0}}} [1+\sigma_{\eta_{c_0}}\times C_2^{3/2}(\xi)]\right)
\nonumber\\
& \times \bigg[\exp\left(-\frac{{\mathbf k}^2_\bot+m^2_c}{8\beta^2_{\eta_c} x(1-x)}\right) \chi(m_c,x,{\mathbf k}_\bot)\bigg],
\end{align}
which will be normalized to the decay constant $f_{\eta_{c_0}}$ similar to Eq.(\ref{Eq:NC}). $f_{\eta_{c_0}}$ should be a positive phenomenological parameter, which characterizes the relative importance of the intrinsic charm components in $\eta (\eta^{\prime})$-meson and can be related to the previously defined decay constant $f^c_{\eta^{\prime}_c}$ by comparing Eqs.(\ref{Eq:eta11}, \ref{Eq:eta12}) with the corresponding expressions in Ref.\cite{Ali:1997nh}, $f_{\eta_{c_0}} =f^c_{\eta^{\prime}_c}/(\cos\theta\tan\theta_c)$. As for the $\eta$-$\eta'$-$\eta_c$ mixing scheme, using the transformation matrix (\ref{eq:fqsc}), we obtain $f_{\eta_{c_0}}=-f^c_{\eta^{\prime}_c}/0.0173$. Similarly, the non-perturbative parameter $f^c_{\eta^{\prime}_c}$ can be determined by comparing with the data. For the convenience of subsequent discussions, we adopt a broader parameter range for our analysis~\footnote{Previously, it has been usually suggested that $f^c_{\eta^{\prime}_c}$ should be negative so as to explain the data on the $\eta-\gamma$ and $\eta'-\gamma$ TFFs. The present transformation matrices (\ref{eq:fqsc}, \ref{eq:fqsgc}) give a natural explanation why it should be negative.}, e.g., $f^c_{\eta^{\prime}_c}=(0, -15, -30, -45, -60)$ MeV~\cite{Feldmann:1997vc, Yuan:1997ts}, which lead to $f_{\eta_{c_0}}=(0, 0.867, 1.734, 2.601, 3.468)~{\rm GeV}$.

To fix the parameters of $\Psi_{\eta_{c_0}}$, we require its first three LCDA moments as $\langle\xi^{\eta_{c_0}}_2\rangle|_{\mu_0}=0.073\pm0.009$, $\langle\xi^{\eta_{c_0}}_4\rangle|_{\mu_0}=0.014\pm0.003$, and $\langle\xi^{\eta_{c_0}}_6\rangle|_{\mu_0}=0.004\pm0.001$~\cite{Zhong:2014fma}. Together with the constraints (\ref{Eq:NC}, \ref{Eq:da}), we then obtain $\beta_{\eta_{c_0}}=4.999~{\rm GeV}$, $\epsilon_{\eta_{c_0}}=2.110$, and $\sigma_{\eta_{c_0}}=-0.096$. We present the TFFs in Fig.~\ref{fig:fetaetac}, in which the curves from top to bottom correspond to $A_{\eta_{c_0}}=0$, $190.016~{\rm GeV^{-1}}$, $380.032~{\rm GeV^{-1}}$, $570.048~{\rm GeV^{-1}}$ and $760.063~{\rm GeV^{-1}}$, respectively. In the low $Q^2$ region, the calculated values of $Q^2 F_{\eta\gamma}(Q^2)$ are consistent with the data within errors, whereas $Q^2 F_{\eta'\gamma}(Q^2)$ exhibits a discrepancy despite following a similar trend of variation with increasing $Q^2$. In the high $Q^2$ region, the intrinsic charm quark contribution is relatively small in the $Q^2F_{\eta\gamma}(Q^2)$ but more significant in  $Q^2F_{\eta^\prime\gamma}(Q^2)$~\footnote{If setting $f_{\eta_{c_0}}=4.335~{\rm GeV}$ or equivalently $f^c_{\eta^{\prime}_c}=-75$ MeV, the predicted $Q^2F_{\eta^\prime\gamma}(Q^2)$ value approaches the central measured value at $Q^2=112 \ {\rm GeV}^2$ reported by BABAR'06~\cite{BaBar:2006ash}.}. Thus it is important to take the intrinsic charm components into consideration.
\begin{figure*}[htb]
	\begin{center}
		\includegraphics[width=0.45\textwidth]{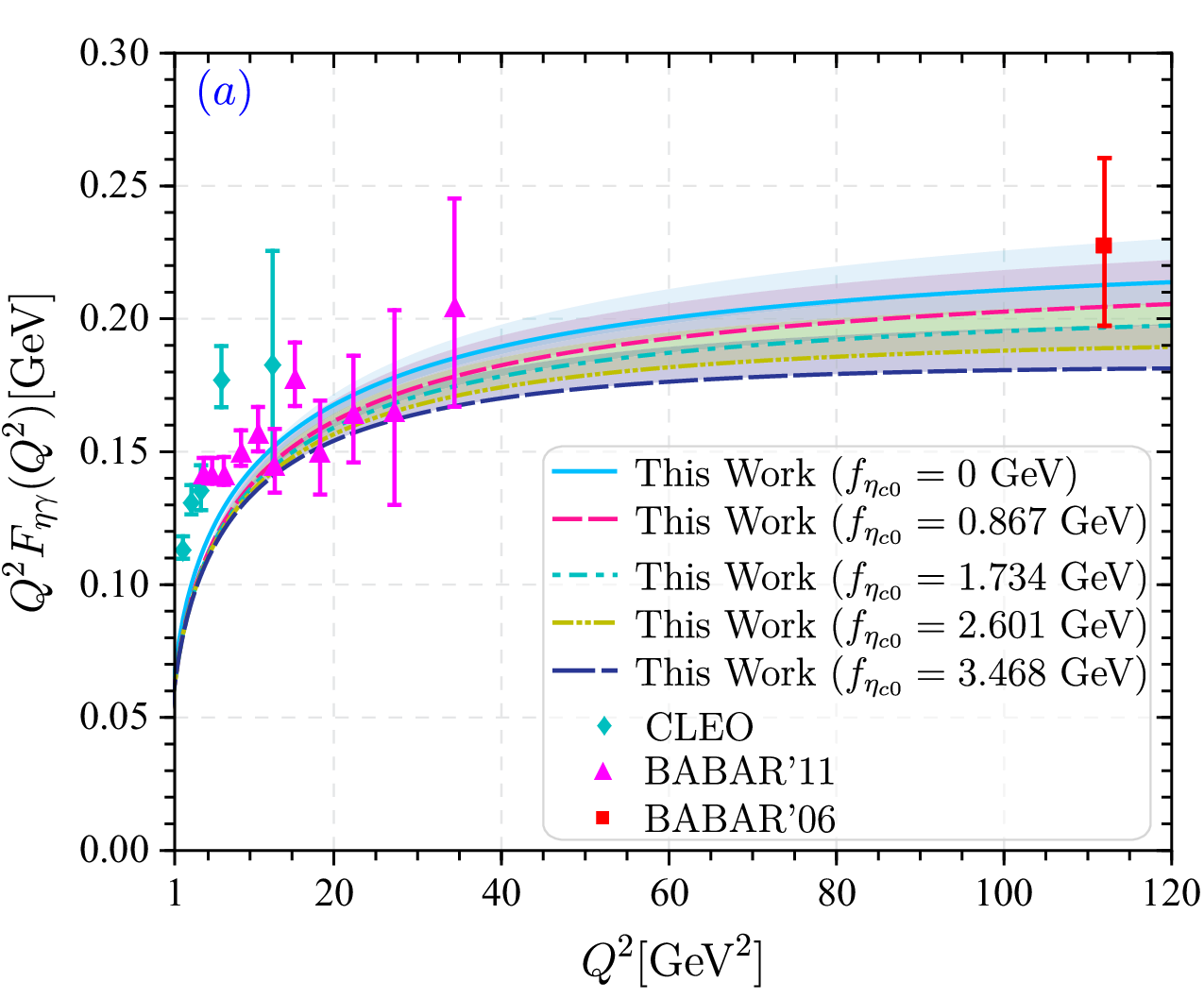}
		\includegraphics[width=0.45\textwidth]{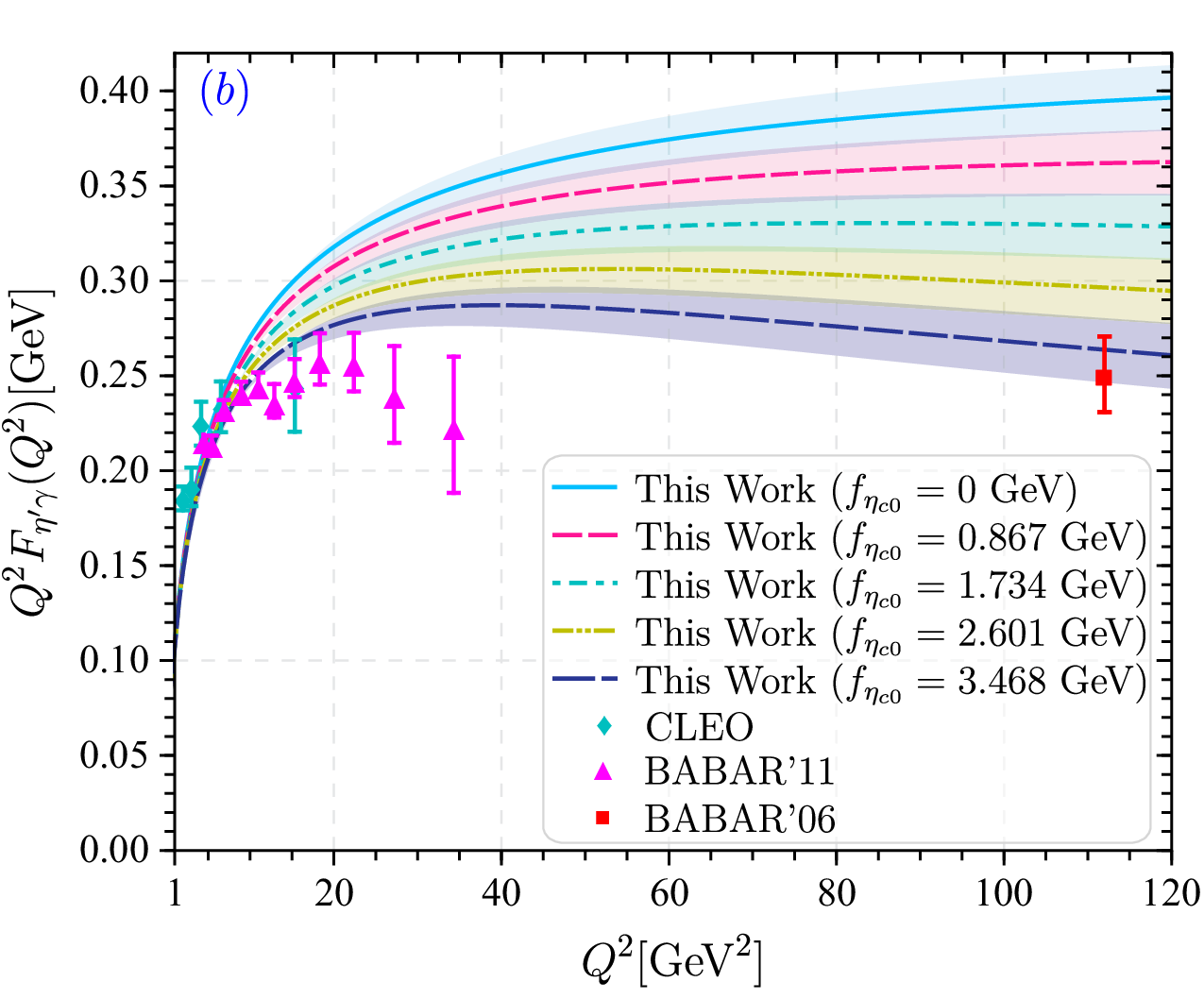}
	\end{center}
	\caption{The $\eta-\gamma$ and $\eta^\prime-\gamma$ TFFs $Q^2F_{\eta\gamma}(Q^2)$ (a) and $Q^2F_{\eta^\prime\gamma}(Q^2)$ (b) are obtained through $\eta-\eta^\prime-G-\eta_c$ mixing mechanism, which the shaded band indicates the uncertainty arising from different choices of light constituent quark masses (e.g. $m_q=0.30\pm0.05~{\rm GeV}$ and $m_s=0.45\pm0.05~{\rm GeV}$). The experimental results from CLEO~\cite{CLEO:1997fho}, BABAR'11~\cite{BaBar:2011nrp} and BABAR'06~\cite{BaBar:2006ash} are also represented.}
	\label{fig:fetacG}
\end{figure*}	

As a step forward, besides the $\eta-\eta^\prime-\eta_c$ mixing scheme, the gluonic components may also have sizable effects to the TFFs. In the following, we discuss this possibility. Using the transformation matrix (\ref{eq:fqsgc}) of the $\eta-\eta'-G-\eta_c$ mixing scheme, we have
\begin{align}
	|\eta\rangle=0.7517|\eta_q\rangle-0.6595 |\eta_s\rangle-0.0005|gg\rangle-0.0049|\eta_{c_0}\rangle,
	\nonumber\\
	|\eta'\rangle=0.6593 |\eta_q\rangle+0.7516|\eta_s\rangle-0.0023|gg\rangle-0.0206 |\eta_{c_0}\rangle. \nonumber
\end{align}
Then the TFFs change to
\begin{eqnarray}
	F_{\eta\gamma}(Q^2) &=& 0.7517\times F_{\eta_q \gamma}(Q^2)-0.6595\times F_{\eta_s \gamma}(Q^2)
	\nonumber\\
	& & -0.0049\times F_{\eta_{c_0}\gamma}(Q^2), \label{Eq:eta21} \\
	F_{\eta'\gamma}(Q^2) &=& 0.6593\times F_{\eta_q \gamma}(Q^2)+0.7516\times F_{\eta_s \gamma}(Q^2)
	\nonumber\\
	& & -0.0206\times F_{\eta_{c_0}\gamma}(Q^2).
	\label{Eq:eta22}
\end{eqnarray}
The gluonic components change the relative importance among the unmixed states $|\eta_q\rangle$, $|\eta_s\rangle$ and $|\eta_{c_0}\rangle$, while its own contributions to the TFFs are negligibly small due to its Feynman diagrams are loop-dominated. Comparing Eqs.(\ref{Eq:eta11}, \ref{Eq:eta12}) with Eqs.(\ref{Eq:eta21}, \ref{Eq:eta22}), we observe that the gluonic components slightly affect the TFFs, which becomes sizable in large $Q^2$ region. Thus it will further improve the precision of the predicted TFFs. Fig.\ref{fig:fetacG} confirms this observation, showing that these two TFFs have similar shapes when compared to those presented in Fig.~\ref{fig:fetaetac} for the same value of $f_{\eta_{c_0}}$. Fig.\ref{fig:fetacG} also shows the dominant uncertainties originating from the variations in light constituent quark masses $m_q$ and $m_s$, with the effect from varying the heavy charm quark mass $m_c$ is very small and has been neglected.

Next, we demonstrate the level of agreement between the predicted TFFs $Q^2 F_{\eta\gamma}(Q^2)$ and $Q^2 F_{\eta^\prime\gamma}(Q^2)$ derived under the $\eta-\eta'-G-\eta_c$ mixing scheme and the available experimental data, with the BABAR dataset~\cite{BaBar:2011nrp, BaBar:2006ash} and the CLEO dataset~\cite{CLEO:1997fho} cited as specific examples. To this end, we employ the $\chi^2/{\rm d.o.f}$ (chi-squared per degree of freedom) metric to quantitatively evaluate the consistency between the theoretical predictions and the experimental measurements, following the methodology outlined in Ref.\cite{ParticleDataGroup:2024cfk}, we have
\begin{displaymath}
	\chi^2/{d.o.f}=\frac{1}{N-1}\sum_{i=1}^{N}\bigg[\frac{Q^2F(Q^2)|_{\rm exp.}-Q^2F(Q^2)|_{\rm the.}}{\delta_i}\bigg]^2.
\end{displaymath}
There is only one free parameter ($A_{\eta_{c_0}}$), which can be determined by $f_{\eta_{c_0}}$. $N$ is the number of data points; it equals $12$ for the BABAR dataset~\cite{BaBar:2011nrp, BaBar:2006ash} and $5$ for the CLEO dataset~\cite{CLEO:1997fho}, respectively. ``exp.'' represents the experimental value and ``the.'' refers to the central value of theoretical prediction. $\delta_i$ represents the combined uncertainty at the $i$-th $Q^2$ point, incorporating both experimental measurement errors and theoretical prediction errors. Experimentally, this encompasses statistical and systematic uncertainties. Theoretically, the main source of uncertainty is the variation in values of the constituent quark masses $m_q$ and $m_s$.

\begin{table}[htb]
	\renewcommand\arraystretch{1.3}
	\center
	\footnotesize
	\caption{$\chi^2/{d.o.f}$ and $p$-values for the TFFs $Q^2 F_{\eta\gamma}(Q^2)$ and $Q^2 F_{\eta^\prime\gamma}(Q^2)$ (the BABAR dataset~\cite{BaBar:2011nrp, BaBar:2006ash}), where typical choices of $f_{\eta_{c_0}}$ (in unit: GeV) are adopted.} \label{Tab:etac:BABAR}
	\begin{tabular}{|c|ccc|}
		\hline
		~~ ~~~   & ~~$f_{\eta_{c_0}}$~~  & ~~$\chi^2/{d.o.f}$~~    ~~& $p$-value~~ \\
\hline
		& $0$       & $0.078$   & $99.9\%$             \\
		& $0.867$       & $0.168$   & $99.8\%$             \\
     	$Q^2F_{\eta\gamma}(Q^2)$ 	& $1.734$      & $0.292$   & $98.7\%$            \\
		& $2.601$     & $0.451$   & $93.3\%$             \\
		& $3.468$     & $0.643$   & $79.2\%$             \\
\hline\hline
		& $0$       & $7.439$   & $\sim 0$             \\
		& $0.867$       & $4.439$   & $\sim 0$             \\
		$Q^2F_{\eta^\prime\gamma}(Q^2)$    & $1.734$     & $2.209$    & $1.15\%$             \\
		& $2.601$     & $0.749$   & $69.1\%$             \\
		& $3.468$     & $0.060$   & $99.9\%$             \\
\hline
	\end{tabular}
\end{table}

\begin{table}[htb]
	\renewcommand\arraystretch{1.3}
	\center
	\footnotesize
	\caption{$\chi^2/{d.o.f}$ and $p$-values for $Q^2 F_{\eta\gamma}(Q^2)$ and $Q^2 F_{\eta^\prime\gamma}(Q^2)$ (the CLEO dataset~\cite{CLEO:1997fho}), where typical choices of $f_{\eta_{c_0}}$ (in unit: GeV) are adopted.} \label{Tab:etac:CLEO}
	\begin{tabular}{|c|ccc|}
		\hline
		~~ ~~~   & ~~$f_{\eta_{c_0}}$~~  & ~~$\chi^2/{d.o.f}$~~    ~~& $p$-value~~ \\
\hline
		& $0$       & $0.295$   & $88.1\%$             \\
		& $0.867$       & $0.323$   & $86.2\%$             \\
     	$Q^2F_{\eta\gamma}(Q^2)$ 	& $1.734$      & $0.352$   & $84.2\%$            \\
		& $2.601$     & $0.383$   & $82.1\%$             \\
		& $3.468$     & $0.414$   & $79.8\%$             \\
\hline\hline
		& $0$       & $1.846$   & $11.7\%$             \\
		& $0.867$       & $1.339$   & $25.3\%$             \\
		$Q^2F_{\eta^\prime\gamma}(Q^2)$    & $1.734$     & $0.913$    & $45.5\%$             \\
		& $2.601$     & $0.568$   & $68.7\%$             \\
		& $3.468$     & $0.305$   & $87.5\%$             \\
\hline
	\end{tabular}
\end{table}

Tables~(\ref{Tab:etac:BABAR}) and (\ref{Tab:etac:CLEO}) present the $\chi^2/{d.o.f}$ values for the TFFs $Q^2F_{\eta\gamma}(Q^2)$ and $Q^2F_{\eta^\prime\gamma}(Q^2)$ derived under the $\eta-\eta'-G-\eta_c$ mixing scheme, where the corresponding $p$-values are also displayed. Table~\ref{Tab:etac:BABAR} shows the results for the BABAR dataset~\cite{BaBar:2011nrp, BaBar:2006ash} and Table~\ref{Tab:etac:CLEO} shows the results for the CLEO dataset~\cite{CLEO:1997fho}. A smaller $\chi^2/{d.o.f}$ value corresponds to a larger $p$-value, indicating better agreement between the predicted values and the data. The $p$-value is given by the follow formula~\cite{ParticleDataGroup:2024cfk}
\begin{displaymath}
p=\int^\infty_{\chi^2/d.o.f}f(t;n_d) dt,
\end{displaymath}
where $f(t;n_d)$ is the probability density function of $\chi^2/{d.o.f}$, $n_d=N-1$ is the appropriate number of degrees of freedom. Tables~(\ref{Tab:etac:BABAR}) and (\ref{Tab:etac:CLEO}) show that the $p$-value of $Q^2 F_{\eta\gamma}(Q^2)$ decreases, while that of $Q^2 F_{\eta'\gamma}(Q^2)$ increases, with the increase of $f_{\eta_{c_0}}$. More explicitly, taking the results of BABAR data an example, the $p$-value of $Q^2F_{\eta\gamma}(Q^2)$ is $99.9\%$ for $f_{\eta_{c_0}}=0$, which changes down to $79.2\%$ for $f_{\eta_{c_0}}=3.468~{\rm GeV}$. While the $p$-value of $Q^2F_{\eta^\prime\gamma}(Q^2)$ is $\sim 0$ for $f_{\eta_{c_0}}=0$, which changes up to $99.9\%$ for $f_{\eta_{c_0}}=3.468~{\rm GeV}$. Thus a proper choice of $f_{\eta_{c_0}}$ can simultaneously explain the data on the TFFs $Q^2F_{\eta\gamma}(Q^2)$ and $Q^2F_{\eta^\prime\gamma}(Q^2)$.

Taking the BABAR dataset as an example, we find that for $f_{\eta_{c_0}}\in[2.601,3.923]$ GeV -- or equivalently $f^c_{\eta^{\prime}_c} \in [-68,-45]$ MeV -- the values of $Q^2 F_{\eta\gamma}(Q^2)$ and $Q^2 F_{\eta^\prime\gamma}(Q^2)$ are consistent with the experimental data within the $1 \sigma$ confidence interval. It is noted that the determined parameter range of $f_{\eta_{c_0}}$ is broader than the decay constant of the $\eta_{c}$ meson (e.g. $f_{\eta_c} \sim 0.566$ GeV under the wavefunction normalization condition adopted herein, as defined in Eq.(\ref{Eq:NC})). This difference is physically reasonable: the decay constant of the intrinsic charmonium component extracted from the $\eta^{(\prime)}$ meson is defined at the $\eta^{(\prime)}$ mass scale. When evolved to the charmonium energy scale via QCD renormalization group running, this quantity receives substantial radiative corrections and is thus significantly enhanced. Consequently, the equivalent intrinsic charmonium decay constant $f_{\eta_{c_0}}$ derived in this work is much larger than the decay constant of the $\eta_c$ meson.

\begin{table}[htb]
    \renewcommand\arraystretch{1.3}
    \center
    \footnotesize
    \caption{$\chi^2/{d.o.f}$ and $p$-values for $Q^2 F_{\eta\gamma}(Q^2)$ and $Q^2 F_{\eta^\prime\gamma}(Q^2)$ under typical choices of $f_{\eta_{c_0}}$ (in unit: GeV) (combined BABAR and CLEO data~\cite{BaBar:2011nrp, BaBar:2006ash, CLEO:1997fho}. } \label{Tab:etac:combined}
    \begin{tabular}{|c|ccc|}
        \hline
        ~~ ~~~   & ~~$f_{\eta_{c_0}}$~~  & ~~$\chi^2/{d.o.f}$~~    ~~& $p$-value~~ \\
        \hline
        & $0$       & $0.078$   & $\sim 100\%$             \\
        & $0.867$       & $0.168$   & $99.9\%$             \\
        $Q^2F_{\eta\gamma}(Q^2)$     & $1.734$      & $0.292$   & $99.7\%$            \\
        & $2.601$     & $0.451$   & $96.8\%$             \\
        & $3.468$     & $0.643$   & $85.1\%$             \\
        \hline\hline
        & $0$       & $7.439$   & $\sim 0$             \\
        & $0.867$       & $4.439$   & $\sim 0$             \\
        $Q^2F_{\eta^\prime\gamma}(Q^2)$    & $1.734$     & $2.209$    & $0.4\%$             \\
        & $2.601$     & $0.749$   & $74.7\%$             \\
        & $3.468$     & $0.060$   & $\sim 100\%$             \\
        \hline
    \end{tabular}
\end{table}

As a final remark, treating the BABAR and CLEO datasets with equal weighting yields a combined result that exhibits better agreement with theoretical predictions. The results are presented in Table~\ref{Tab:etac:combined}, which shows the $\chi^2/{d.o.f}$ values for the combined dataset are identical to those obtained using the BABAR data alone, while the $p$-values are significantly improved. It is therefore anticipated that our current predictions will be further validated by more precise experimental data, such as those from the BELLE II collaboration.

\section{Summary}\label{sec:summary}

In this study, we employed the light-cone pQCD to make an improved investigation on the $\gamma\gamma^*\to \eta^{(\prime)}$ TFFs $Q^2F_{\eta\gamma}(Q^2)$ and $Q^2F_{\eta'\gamma}(Q^2)$, incorporating transverse momentum corrections. Our analysis accounts for both the $\eta$-$\eta'$ mixing effects and the contributions beyond the leading Fock state. Specifically, we adopted the quark-flavor mixing scheme for the calculation of the TFFs and used a parameterization to estimate higher Fock-state contributions. Furthermore, we applied two distinct models to describe the wavefunctions of the $\eta_q$ and $\eta_s$ mesons. The behaviors of leading-twist LCDA resulting from LCHO models are shown in Fig. \ref{fig:DA}, where it can be seen that they all exhibit a unimodal form.

To estimate the theoretical uncertainties, we examined the impact of the dominant error source -- the constituent quark masses -- on the TFFs and found that this effect is relatively minor for both $Q^2F_{\eta\gamma}(Q^2)$ and $Q^2F_{\eta^\prime\gamma}(Q^2)$. A combined analysis of Figs.~\ref{fig:fetaetac} and \ref{fig:fetacG} indicates that both TFFs agree with the experimental data in the low-$Q^2$ region with high probability, supporting their universal behavior at low energies. However, at larger $Q^2$ values, $Q^2F_{\eta^\prime\gamma}(Q^2)$, which is more sensitive to the charm quark component shows significantly better consistency with the BABAR'06 data. This suggests that by taking the intrinsic charm and the gluonic components into consideration, a more accurate description of the TFFs can be achieved. And a detailed analysis on the TFFs has been done under the $\eta$-$\eta^\prime$-$G$-$\eta_c$ mixing mechanism, which confirms this observation.

To better characterize the consistency between the TFFs and experimental data, we also calculated their corresponding $\chi^2/d.o.f$ and $p$-values. For $f_{\eta_{c_0}}$ varying within the range $[2.601,3.923]$ GeV, the predictions for both $Q^2F_{\eta\gamma}(Q^2)$ and $Q^2F_{\eta^\prime\gamma}(Q^2)$ are consistent with the earlier BABAR dataset -- comprising 12 data points -- within the $1 \sigma$ confidence interval. These comparisons provide valuable experimental benchmarks for testing and advancing QCD theory. If confirmed by future experiments, these findings could either hint at new physics or deepen our understanding of QCD itself, while also illustrating that subtle differences in the internal structure of seemingly similar mesons can lead to distinct behaviors in high-energy scattering processes. By combining the given BABAR and CLEO datasets with equal weighting, Table~\ref{Tab:etac:combined} shows that improved agreement with theoretical predictions can be achieved. The updated Belle II experiment and the SuperKEKB accelerator~\cite{Belle-II:2010dht, Akai:2018mbz} are now in operation, which enables more precise measurements of the TFFs and other related observables. We anticipate that the present work will serve as a theoretical foundation for these future experimental studies.

\section{Acknowledgments}

This work was supported in part by the National Natural Science Foundation of China under Grant No.12265010, No.12265009, No.12575080 and No.12547101, the Project of Guizhou Provincial Department of Science and Technology under Grant No.MS[2025]219, No.CXTD[2025]030 and No.ZK[2023]024, the Graduate Research and Innovation Foundation of Chongqing, China under Grant No.ydstd1912.   \\

\appendix

\section*{Appendix: The transformation matrixes}
\label{app:transformation matrix}

The transformation matrices involved in Eqs.\eqref{eq:qsc} and \eqref{eq:qsgc} are
\begin{widetext}
\begin{eqnarray}
	U_{3}(\theta)=\left( \begin{array}{ccc}
		\cos\theta & -\sin\theta & 0 \\
		\sin\theta & \cos\theta &0 \\
		0 &0 &1
	\end{array} \right), ~~~
	U_{1}(\theta_c)=\left( \begin{array}{ccc}
		1 &0 &0 \\
		0 &\cos\theta_c & \sin\theta_c \\
		0 &-\sin\theta_c & \cos\theta_c
	\end{array} \right),
\end{eqnarray}
\begin{eqnarray}
	& &U_{34}(\theta)=\left( \begin{array}{cccc}
		\cos\theta & -\sin\theta & 0 &0\\
		\sin\theta & \cos\theta &0 &0\\
		0 &0 &1 &0\\
		0 &0 &0 &1
	\end{array} \right), ~~~
	U_{14}(\phi_g)=\left( \begin{array}{cccc}
		1 &0 &0 &0\\
		0 &\cos\phi_g & \sin\phi_g &0 \\
		0 &-\sin\phi_g & \cos\phi_g &0\\
		0 &0 &0 &1
	\end{array} \right), \\
	& &U_{13}(\phi_c)=\left( \begin{array}{cccc}
		1 &0 &0 &0\\
		0 &\cos\phi_c & 0 &\sin\phi_c \\
		0 &0 &1 &0\\
		0 &-\sin\phi_c & 0 &\cos\phi_c
	\end{array} \right), ~~~
	U_{12}(\theta_g)=\left( \begin{array}{cccc}
		1 &0 &0 &0\\
		0 &1 &0 &0\\
		0 &0 &\cos\theta_g & \sin\theta_g  \\
		0 &0 &-\sin\theta_g & \cos\theta_g
	\end{array} \right),
\end{eqnarray}
\end{widetext}

\end{document}